\documentclass[twocolumn,nofootinbib,aps,tightenlines,preprintnumbers,notitlepage,longbibliography,superscriptaddress]{revtex4-1}
\usepackage{graphicx}
\usepackage{amsmath}
\usepackage{amsfonts}
\usepackage{comment}
\usepackage[colorlinks=true,citecolor=blue,urlcolor=cyan,linkcolor=blue]{hyperref}

\usepackage{amsmath}
\usepackage{amsfonts}

\usepackage{bm}
\usepackage{tabularx}
\newcolumntype{C}{>{\centering\arraybackslash}X}
\newcolumntype{R}{>{\raggedleft\arraybackslash}X}

\usepackage{romannum}
\usepackage[usenames, dvipsnames]{color}
\usepackage[normalem]{ulem}
\usepackage{xcolor}

\newcommand{\be}{\begin{eqnarray}}

\newcommand{\ee}{\end{eqnarray}}

\definecolor{colorA}{HTML}{1E90FF}
\definecolor{colorB}{HTML}{228B22}
\definecolor{colorC}{HTML}{FF7F00}
\definecolor{colorD}{HTML}{4B0082}
\definecolor{colorE}{HTML}{B22222}

\definecolor{lgreen}{HTML}{32CD32}
\definecolor{lgray}{HTML}{D3D3D3}
\definecolor{dblue}{HTML}{1E90FF}
\definecolor{dblue}{HTML}{1E90FF}
\definecolor{orange}{HTML}{FF4500}
\definecolor{indigo}{HTML}{4B0082}

\definecolor{teal}{HTML}{008080}
\definecolor{firebrick}{HTML}{B22222}
\definecolor{salmon}{HTML}{FA8072}
\definecolor{darkgreen}{HTML}{006400}

\newcommand{\jhu}{William H. Miller III Department of Physics and Astronomy, Johns Hopkins University, Baltimore, MD 21218, USA}

\newcommand{\nyu}{Center for Cosmology and Particle Physics, Department of Physics, New York University, New York, NY 10003, USA}

\graphicspath{ {plots/} }

\begin{document}

\pagenumbering{arabic}

\title{Probing light relics through cosmic dawn}

\author{Nanoom~Lee}
\email{nanoom.lee@nyu.edu}
\affiliation{\nyu}

\author{Selim~C.~Hotinli}
\email{selimcanhotinli@gmail.com}
\affiliation{\jhu}

\begin{abstract}

We explore the prospects of upcoming 21{-}cm surveys of cosmic dawn ($12\lesssim \!z\lesssim\!30$) to provide cosmological information on top of upcoming cosmic microwave background (CMB) and large-scale structure surveys, such as CMB-S4, Simons Observatory (SO) and DESI. We focus on {the effective number of relativistic species $N_{\rm eff}$} which is a promising observable for probing beyond the Standard Model theories. We show including upcoming 21-cm surveys such as the Square Kilometre Array (SKA) can allow probing a wide range of models for light particles at $2\sigma$ level achieving $2\sigma(N_{\rm eff})=0.034$ with CMB-S4, for example. Taking into account the degeneracy between $N_{\rm eff}$ and primordial helium fraction $Y_p$, one can achieve improvements in sensitivities to cosmological parameters, in particular, by more than a factor of 2 for $N_{\rm eff}$ and dark matter fractional energy density $\omega_c$.

\end{abstract}

\maketitle

\section{Introduction}

Data from current and future cosmological surveys such as CMB-S4 \cite{CMB-S4:2016ple, Abazajian:2019eic, Abazajian:2022nyh}, Simons Observatory \cite{Ade:2018sbj,SimonsObservatory:2019qwx}, CMB-HD \cite{Sehgal:2019ewc,CMB-HD:2022bsz}, DESI \cite{DESI:2016fyo,DESI:2019jxc}, the Rubin Observatory Legacy Survey of Space and Time (LSST) \cite{LSSTDarkEnergyScience:2012kar, LSSTDarkEnergyScience:2018jkl,LSST:2008ijt}, the Nancy Grace Roman Space telescope \cite{Spergel:2015sza}, and the Euclid satellite \cite{EUCLID:2011zbd} will provide a wealth of new information on our Universe, improving the sensitivity to a wide range of cosmological parameters. Among these, the effective number of relativistic species $N_{\rm eff}$ is of particular interest amongst cosmology and particle physics communities as it incorporates any light relics [particles from the early Universe that are relativistic at recombination ($m<1$eV)] that make their contributions to the total energy density of radiation, providing a window into probing very high energy physics. 

{Appealingly, broad classes of light particles beyond the Standard Model (SM) can be simply described by three values of $\Delta N_{\rm eff}$ (the difference between the observed $N_{\rm eff}$ and its SM prediction) depending on their spins, given a freeze-out temperature, $T_F$ (see the left panel of Fig.~\ref{fig:forecast}) \cite{CMB-S4:2016ple,Abazajian:2019eic}. {Consequently, a measurement sensitivity} reaching $\Delta N_{\rm eff}\lesssim 0.047$ or $0.027$ excludes light particles with spin 1, 1/2 or spin 0, respectively. Upcoming CMB-S4 survey is expected to marginally reach this level of sensitivity, for example, and forecasted to achieve $\sigma(N_{\rm eff})\approx0.03$ \cite{CMB-S4:2016ple,Abazajian:2019eic}.} In this work, we show how measurements of 21-cm line intensity from cosmic dawn can improve the $N_{\rm eff}$ measurement significantly.

The relativistic species affect CMB anisotropy spectra in two ways. They affect the expansion rate during radiation domination, impacting the diffusion damping scale, and thus the amplitude of CMB spectra on small scales. Since the fluctuations of relativistic species travel faster than the sound speed of the baryon-photon plasma, they also lead to a characteristic phase shift of the baryonic acoustic oscillations (BAOs)~\cite{Bashinsky:2003tk, Hou:2011ec, Follin:2015hya}. Importantly, it is difficult to reproduce the effect of phase shifts in the absence of free streaming radiation~\citep{Bashinsky:2002vx,Eisenstein:2006nj,Baumann:2017lmt,Baumann:2017gkg,Baumann:2019keh}. This makes high-precision measurements of BAOs a promising route to improving the sensitivity to $N_{\rm eff}$.

The cosmic dawn, or the intermediate epoch in time when the first stars were formed ($12\lesssim z\lesssim30$), holds rich cosmological information complementary to CMB and large-scale structure (LSS). While modeling this signal heavily relies on modeling of astrophysical effects, large-scale spatial correlations of 21-cm temperature brightness are very sensitive to cosmology, largely through the imprint of the relative-velocity differences between cold dark matter (CDM) and baryons, which play an important role on regulating the collapse of baryonic objects at early times \cite{Tseliakhovich:2010bj}, leading to a distinct velocity acoustic oscillation (VAO) signature~\cite{Munoz:2019rhi}. Note that we focus on the cosmic dawn in this work as the VAO signature which is sensitive to $N_{\rm eff}$ gets weaker at lower redshifts \cite{Machacek:2000us,Fialkov:2012su,Visbal:2014fta}.

Since the relative velocity differences between CDM and baryons are sourced during recombination, VAOs are analogous to BAOs, providing a new standard ruler at high redshifts~\cite{Munoz:2019fkt}. However, the high sensitivity of newly-forming baryonic objects to the baryon-CDM relative velocity results in VAOs to have an $\mathcal{O}(1)$ effect on the 21-cm spatial correlations during this time, an effect much larger than the BAO imprint on the LSS~\citep[see~e.g.][for applications]{Hotinli:2021vxg,Hotinli:2021xln}. Similar to BAOs, the VAO signature is also sensitive to $\Delta N_{\rm eff}$ as well as other cosmological parameters with different degeneracies that can further enhance the sensitivity to cosmological parameters upon its joint analysis with other cosmological surveys. 

An important factor relevant to the measurement sensitivity of $N_{\rm eff}$ is the primordial helium fraction $Y_p$ which affects the free electron fraction prior to recombination and leads to a degenerate effect on the diffusion damping of the CMB spectra (see, for example, Fig.~3 of Ref.~\cite{Hou:2011ec}), closely related to the ratio $\theta_d/\theta_s$, which is the ratio between the angular scales of diffusion damping and sound horizon scale \cite{Hou:2011ec}. While the big bang nucleosynthesis (BBN) consistency provides a clear relation between $N_{\rm eff}$ and $Y_p$ through its sensitivity to the expansion rate of the Universe, here we also consider testing this consistency relation by setting $Y_p$ to be a free parameter, allowing beyond standard BBN models that probe deviations from the standard thermal history of the Universe and the SM of particle physics~\cite{CMB-S4:2016ple} or different astrophysical scenarios~\citep[e.g.][]{Hotinli:2022jna,Hotinli:2022jnt}.

In what follows, we model 21-cm power spectra by fitting semi-numerical simulations from \texttt{21cmvFAST}\footnote{\href{https://github.com/JulianBMunoz/21cmvFAST}{github.com/JulianBMunoz/21cmvFAST}} \cite{Munoz:2019rhi} with a simple fitting form, and perform ensemble-statistics analysis assuming a SKA1-low--like 21-cm survey, jointly with CMB-S4 and DESI BAO. We show that upcoming 21-cm surveys will indeed provide significant additional sensitivity for constraining cosmological parameters, {particularly allowing exclusion of a wide range of models for light particles with spin 1 and 1/2 at $2\sigma$ level}. Varying both $N_{\rm eff}$ and $Y_p$, we find that 21-cm surveys will improve our ability to constrain cosmology, for example, by more than a factor of 2 for $N_{\rm eff}$ and $\omega_c$.

\section{Cosmic Dawn}

The spin-flip transition between the two hyperfine levels of ground state hydrogen creates a 21-cm emission or absorption line, which gets redshifted as the Universe expands. Observation of the redshifted 21-cm lines at different frequencies allows probing a range of epochs between recombination and today, including the epoch of cosmic dawn. {High-significance measurements of cosmic dawn signals are attainable and upcoming~\citep[see~e.g.][]{Bowman:2018yin, Singh:2017syr, 2019JAI.....850004P,DAPPER2021}, with measurements already performed by the EDGES collaboration~\citep{Bowman:2018yin}, and several 21-cm fluctuations surveys on the way, including the ongoing HERA~\cite{DeBoer:2016tnn} and upcoming SKA1-low~\citep{SKA:2018ckk,2019arXiv191212699B}.}

Modeling 21-cm fluctuations during this epoch can be complicated due to astrophysical processes. Here we use \texttt{21cmvFAST} \cite{Munoz:2019rhi}, a modified version of \texttt{21cmFAST} \cite{Mesinger:2010ne, Greig:2015qca} that takes into account the effect of baryon-CDM relative velocities\footnote{This modification now is implemented in the current version of \texttt{21cmFAST} \cite{Munoz:2021psm} (\href{https://github.com/21cmfast/21cmFAST}{github.com/21cmfast/21cmFAST}).}. \texttt{21cmvFAST} incorporates the relative velocities in its initial conditions, which affect the amount of stellar formation and modify the 21-cm power spectra during Lyman-$\alpha$ coupling era (LCE) \cite{Dalal:2010yt} and during the epoch of heating (EoH) \cite{Visbal:2012aw,McQuinn:2012rt} due to X-ray heating \cite{Pritchard:2006sq,Mesinger:2012ys}, which have a large impact on 21-cm signal from $12\lesssim z \lesssim 30$ \cite{Dalal:2010yt, Visbal:2012aw, McQuinn:2012rt, Fialkov:2012su, Fialkov:2013uwm, Munoz:2019rhi}, in a way dependent on astrophysical model, such as the Lyman-Werner (LW) feedback \cite{Wise:2007nb,Machacek:2000us,Visbal:2014fta}. In what follows, we use predictions from \texttt{21cmvFAST} for a range of feedback levels defined in Ref.~\cite{Munoz:2019rhi}. That is, we consider three feedback levels (low, regular, and high). The low and regular feedback levels are defined to increase the mass of cooling halos $M_{\rm cool}$ by a factor of $[1+B(F_{\rm LW})^\beta]$ with $(B,\beta) = (4,0.47)$, and $(7,0.47)$ respectively, where $F_{\rm LW}$ is the LW flux. The high feedback level is defined to increase the minimum virial velocity that a halo requires to cool through molecular hydrogen by a factor of $[1+B(F_{\rm LW})^\beta]^{1/3}$ with $(B,\beta) = (7,0.47)$. In these models, the average LW flux from the strong-feedback case in Ref.~\cite{Fialkov:2012su} is taken. The different feedback levels take into account the effect of relative velocity on the minimum virial velocity, modeled as $V_{\rm cool}(z,v_{bc}) = [(V_{\rm cool}^{0})^2+\alpha^2v_{bc}^2(z)]^{1/2}$, differently with $\alpha=4$ for regular and high feedback levels and $\alpha=6$ for low feedback level. Note that the fraction of gas in stars depends on the LW feedback as it is parameterized as $f_*(M) \propto \log (M/M_{\rm cool})$ following Ref.~\cite{Fialkov:2012su}.

\begin{figure*}[!t]
\centering
\includegraphics[width=.49\linewidth, trim= 10 15 10 0]{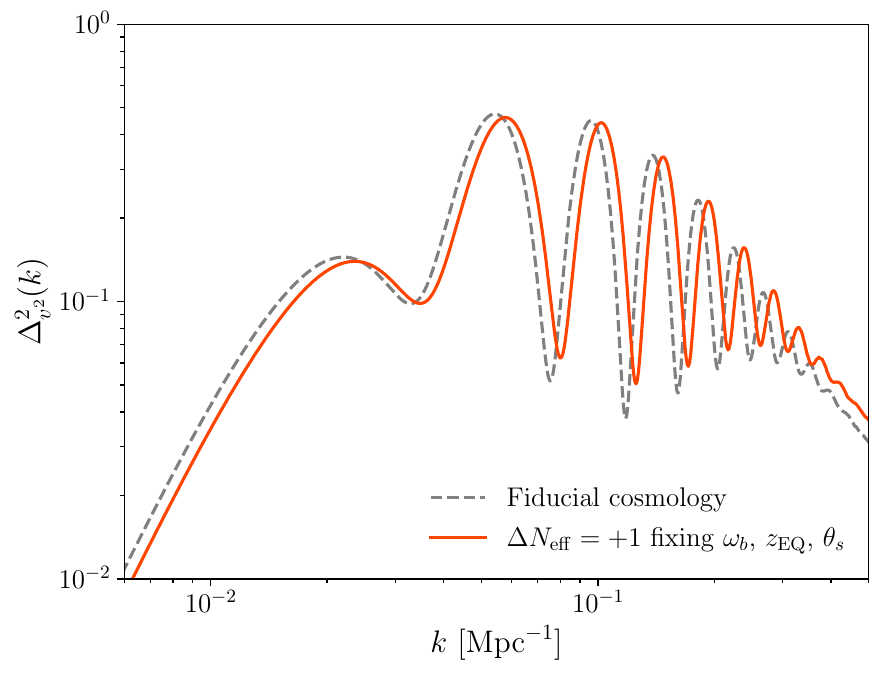}
\includegraphics[width=.495\linewidth, trim= 0 15 27 0]{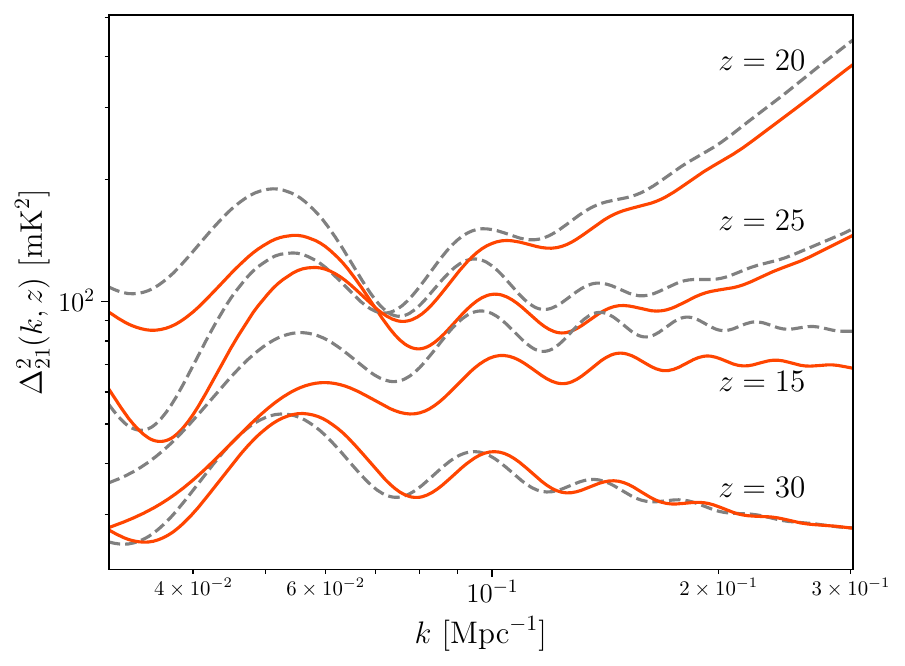}
\caption{Left: Dimensionless power spectrum of $\delta_{v^2}\equiv\sqrt{3/2}\left(v_{bc}^2/v^2_{\rm rms}-1\right)$, where $v_{bc}$ is the relative velocity between CDM and baryons and $v_{\rm rms}\approx 29\;{\rm km/s}$ is its root-mean-square. The power spectrum is calculated following the method developed in Ref.~\cite{Ali-Haimoud:2013hpa}. The transfer functions for CDM and baryon velocity divergence at the baryon drag epoch ($z\approx 1060$) are extracted from \texttt{CLASS}\footnote{\href{https://github.com/lesgourg/class_public}{github.com/lesgourg/class\_public}} \cite{Blas:2011rf} inherently running \texttt{HYREC-2}\footnote{\href{https://github.com/nanoomlee/HYREC-2}{github.com/nanoomlee/HYREC-2}} \cite{Ali-Haimoud:2010tlj,Ali-Haimoud:2010hou,Lee:2020obi} for cosmological recombination calculations. Right: 21-cm power spectra at $z=15,20,25,30$ from Eq.~\eqref{eq:fit}. In both panels, grey dashed curves are with fiducial cosmology, and red curves with $\Delta N_{\rm eff}=+1$ while fixing $\omega_b$, $z_{\rm EQ}$ (the redshift of matter-radiation equality), and $\theta_s$ following Ref.~\cite{Hou:2011ec}. A change in $N_{\rm eff}$ shifts peaks of VAOs.}
\label{fig:D2v2}
\end{figure*}

\section{Forecasts}

We forecast with the goal of assessing the added benefit from 21-cm cosmic dawn on improving the sensitivity to cosmological parameters. To this end, we primarily consider CMB-S4 \cite{CMB-S4:2016ple} and DESI \cite{DESI:2016fyo}, in conjuction with the upcoming 21-cm data from the low-frequency component of SKA (SKA1-low) which we describe next. We calculate the information matrix for CMB-S4 and DESI based on survey specifications provided in Refs.~\cite{CMB-S4:2016ple,DESI:2016fyo}\footnote{We model the CMB at each frequency channel including the pink and white noise components matching anticipated specifications of next-generation CMB experiments, ranging from the upcoming SO to the future CMB-S4 following Table~2 of Ref.~\citep{Hotinli:2021hih}, and use public codes \href{https://github.com/ctrendafilova/FisherLens}{github.com/ctrendafilova/FisherLens} and \href{https://github.com/selimhotinli/class_delens}{github.com/selimhotinli/class\_delens} when generating CMB and DESI BAO information matrices~\citep{Hotinli:2021umk}.}.

We model 21-cm power spectra following Refs.~\cite{Munoz:2019rhi,Munoz:2019fkt}. We take 21-cm power spectra from \texttt{21cmvFAST} and fit a 4th order polynomial in following form,
\be
\Delta_{21}^2(k,z) &=& \mathcal{P}_4(k,z) + A_{\rm vel}(z)\Delta_{v^2}^2(k)\Big|W_i(k,z)\Big|^2,\label{eq:fit}\\
\ln \mathcal{P}_4(k,z) &\equiv&  \sum_{i=0}^4 c_i(z) \big[\log(k)\big]^i\label{eq:poly}
\ee
where $W_i(k,z)=W_X(k,z)$ during EoH and $W_\alpha(k,z)$ during LCE (see Ref.~\cite{Munoz:2019rhi} and references therein for details on astrophysical effects during EoH and LCE, and expressions for the window functions $W_i$). Here $\Delta^2_{v^2}(k)$, shown in the left panel of Fig.~\ref{fig:D2v2}, is the dimensionless power spectrum of $\delta_{v^2}\equiv \sqrt{3/2}(v^2_{bc}/v^2_{\rm rms}-1)$, where $v_{bc}$ is the relative velocity between CDM and baryons, $v_{\rm rms}\approx 29\;\text{km/s}$ is its root-mean-square. In addition to the cosmological parameters, we have six more parameters $\{c_0(z),\ldots,c_4(z),A_{\rm vel}(z)\}$ for each redshift bin, which we marginalize over. We assume that this accounts for most of the dependence on uncertain astrophysics and modelling errors~\cite[latter discussed in e.g.][]{Greig:2015qca}. We take five redshift bins centered at $z = \{12,14,16,18,24\}$ with widths $\Delta z=3$ for $z=24$ and $\Delta z=2$ for the rest. We use the angular spectra~\cite{Zahn:2005ap}
\be
C^{21}_{\ell}(k_\parallel,z) = \frac{(2\pi^2/k^3)\Delta^2_{21}\left(\sqrt{k_\perp^2 + k_\parallel^2},z\right)}{\chi^2(z)\Delta\chi(z)},
\ee
where $k=\sqrt{k_\perp^2+k_\parallel^2}$, $k_\perp = \ell/\chi$, and $k_\parallel = 2\pi j/\Delta \chi$. Here, $\chi$ is the comoving distance to the redshift-bin center, and $\Delta \chi$ is the redshift-bin width. The range of integer $j$ is to be set by $k_{\parallel,{\rm min}}$ and $k_{\parallel,{\rm max}}$ at each redshift. We fix $k_{\parallel,{\rm max}}=0.3\;\text{Mpc}^{-1}$ throughout the forecasts and we define $k_{\parallel,{\rm min}}$ as
\be
k_{\parallel,{\rm min}}=a + b\;k_\perp,
\label{eq:kmin}
\ee
to reflect the realistic form of foreground wedge suffered by interferometric 21-cm experiments~\citep{Liu:2014bba,Liu:2014yxa}. We vary $a$ and $b$ within the ranges $a \in (0.01,0.2)\;\text{Mpc}^{-1}$ and $b \in (0.25,7)$. We also present results considering two cases of foregrounds \{Foreground \Romannum{1}, Foreground \Romannum{2}\} corresponding to $(a,b) = (0.05\;\text{Mpc}^{-1}, 2)$ and $(0.01\;\text{Mpc}^{-1},1)$.

We set the observational noise on the 21-cm power spectrum using the interferometer thermal noise expression from Ref.~\cite{Romeo:2017zwt} as
\be
C^N_\ell(k_\parallel) &=& \left[\frac{\lambda^2}{A_e} F\left(\frac{\nu}{\nu_c}\right)\right]^2 \frac{T_{\rm sys}^2(\nu)}{N_{\rm pol}B t_0} \frac{1}{n(\bm{\ell}/2\pi,\nu)},\\
F(x) &\equiv& \begin{cases} 1,& x\leq1\\x^2,& x>1\end{cases},
\ee
where $A_e=925\;\text{m}^2$ is the effective receiving area of a single station, $\nu_c=110\;{\rm MHz}$ is the ``critical frequency" above which the effective receiving area receives a multiplicative correction of $(\nu_c/\nu)^2$, $N_{\rm pol}=2$ is the number of polarizations per receiver, $B$ is the observing bandwidth to be determined by the width of redshift bin, and $t_0=2000\;{\rm hrs}$ is the total observing time. $T_{\rm sys}$ which is the fundamental source of thermal noise can be approximated as $T_{\rm sys}(\nu) = 40\;{\rm K} + 66 [{\nu}/{(300\;{\rm MHz})}]^{-2.55}\;{\rm K}$.
The time-averaged number density of baselines in $uv$ plane, $n(\bm{u},\nu)\approx \int d^2\bm{x} \mathcal{P}(\bm{x}+\lambda \bm{u})\mathcal{P}(\bm{x})$, where $\bm{u}=\bm{\ell}/2\pi$ and $\mathcal{P}(\bm{x})$ is the radial profile of the antenna distribution on the ground \cite{Foreman:2018gnv} satisfies $n(\bm{\ell}/2\pi, \nu) \simeq 0.5\exp[-9.23\times 10^{-10}\ell^2(1+z)^2]$. Note that this function must be normalized to equal the number of receiver pairs when integrated over the upper-half-$uv$-plane as $\int_{\rm UHP} d^2\bm{u}\,n(\bm{u},\nu) = {N_{\rm rec}(N_{\rm rec}-1)}/{2}$, where we take $N_{\rm rec}=433$ for SKA1-low. 

We calculate the information matrix for SKA1-low with cosmological parameters and fitting coefficients (six for each redshift bin, hence total of thirty for five redshift bins we consider) as
\be
F_{\alpha\beta} = \sum_{z,j,\ell} \frac{f_{\rm sky}(2\ell+1) \partial_\alpha C_\ell^{21}(k_\parallel,z) \partial_\beta C_\ell^{21}(k_\parallel,z)}{2\Big[C_{\ell}^{21}(k_\parallel,z) + C_{\ell}^N(k_\parallel,z)\Big]^2},
\ee
where we set $f_{\rm sky}=5\times10^{-3}$, $\alpha$/$\beta$ indicate considered parameters, and $\partial_\alpha  \equiv \partial/\partial \alpha$. Note that among various feedback levels defined in \texttt{21cmvFAST}, we take `regular feedback' as our fiducial choice when varying foreground settings [$a$ and $b$ in Eq.~\eqref{eq:kmin}]. We also explore the impact of different feedback levels with two fixed foreground settings (Foreground \Romannum{1} and \Romannum{2}). Note that with regular feedback, we get total signal-to-noise ratio over all redshift bins, $\text{SNR}= \big\{\sum_{z,j,\ell}  f_{\rm sky} (2\ell+1)\big[C_{\ell}^{21}(k_\parallel,z)\big]^2 / 2\big[C_{\ell}^{21}(k_\parallel,z) + C_{\ell}^N(k_\parallel,z)\big]^2\big\}^{1/2}=\{183,\;258\}$ for Foreground \Romannum{1} and \Romannum{2}.

\begin{figure*}[!ht]
\centering\includegraphics[width=\linewidth, trim= 10 10 10 10]{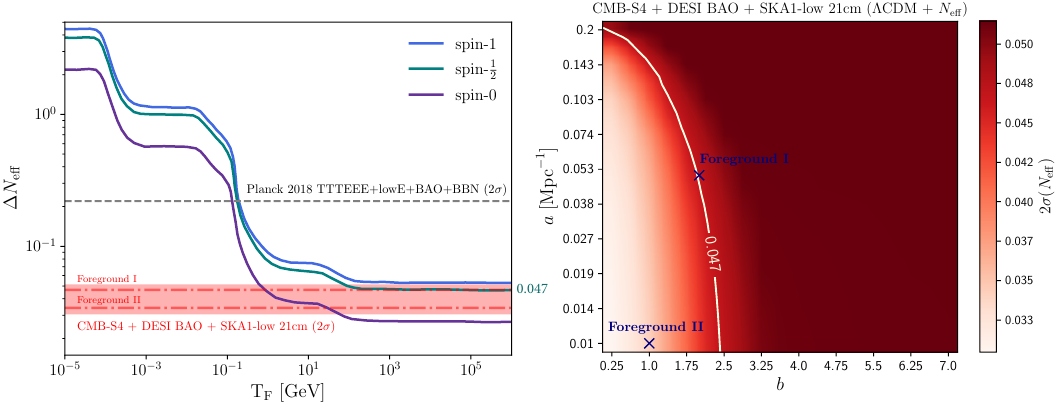}
\caption{Both panels show $2\sigma$ sensitivities to $N_{\rm eff}$ for $\Lambda$CDM + $N_{\rm eff}$ ($Y_p$ is varied to be consistent with BBN). Left: The red band indicates $2\sigma$ sensitivities to $N_{\rm eff}$ when varying foreground settings $k_{\parallel,{\rm min}} = a + b\;k_\perp$ with $a\in (0.01,0.2)\;\text{Mpc}^{-1}$ and $b\in(0.25,7)$. The upper boundary of the red band indicates $2\sigma$ sensitivity to $N_{\rm eff}$ of CMB-S4 + DESI BAO forecast. Forecasts with two foreground settings defined in the main text, which are Foreground \Romannum{1} $(a,b)=(0.05\;\text{Mpc}^{-1},2)$ and Foreground \Romannum{2} $(a,b)=(0.01\;\text{Mpc}^{-1},1)$ are shown as red dot-dashed lines. Right: $2\sigma$ sensitivities to $N_{\rm eff}$ as a function of $a$ and $b$. The white level-contour corresponds to when $2\sigma$ sensitivities are identical to the extra contribution from spin 1/2 particles to $N_{\rm eff}$ with high freeze-out temperature $T_F \gtrsim 10^3\;\text{GeV}$, which is $\Delta N_{\rm ef}=0.047$ (see green curve in the left panel).}
\label{fig:forecast}
\end{figure*}

We present our forecasts on $N_{\rm eff}$ for the standard 6-parameter cosmological model $\Lambda$CDM with the addition of $N_{\rm eff}$ in Fig.~\ref{fig:forecast}\footnote{We take \textit{Planck} $\Lambda$CDM best-fits \cite{Planck:2018vyg} as our fiducial cosmology, and set the effective number of relativistic species as $N_\mathrm{eff}=3.046$.}. 
Here, we have chosen the regular feedback setting in $\texttt{21cmvFAST}$ and vary foregrounds [or equivalently $k_{\parallel,{\rm min}}$ by varying $a$ and $b$ in Eq.~\eqref{eq:kmin}]. In the left panel, the three colored solid curves indicate the contributions to $\Delta N_{\rm eff}$ from light particles with spin 1, 1/2, and 0 as a function of their freeze-out temperature $T_F$. The grey dashed line is Planck 2018 TTTEEE+lowE+BAO+BBN $2\sigma$ constraint on $N_{\rm eff}$ \cite{Planck:2018vyg}; showing that it can only exclude such particles with relatively low freeze-out temperature $T_F \lesssim 10^{-1}\;\text{GeV}$. The red band indicates the range of $2\sigma$ sensitivity to $N_{\rm eff}$ achievable with a SKA1-low--like survey, together with CMB-S4 and DESI BAO, depending on a wide range of foreground settings. The sensitivity of CMB-S4 + DESI BAO, which corresponds to the upper boundary of the red band, only marginally probes models for spin-1 light particles at the $2\sigma$ level. {The extension of the red band to lower values of $\Delta N_{\rm eff}$ indicates the potential of upcoming 21-cm surveys, which can allow probing both spin 1 and 1/2 light particles. The dot-dashed lines further indicate the sensitivities corresponding to two foreground settings, \{Foreground \Romannum{1}, Foreground \Romannum{2}\}. 

The right panel of Fig.~\ref{fig:forecast} demonstrates the sensitivity of 21-cm measurements to astrophysical foregrounds. The white contour line corresponds to the sensitivity required to probe spin 1/2 particles at $\Delta N_{\rm eff}=0.047$ with $2\sigma$ confidence. Note that here $b\approx1$ corresponds to an optimistic scenario of a `primary beam wedge' as often assumed in literature, while for redshifts we consider the `horizon limit' gives $b(z)\!>\!4$. This suggests achieving $2\sigma(N_{\rm eff})\!=\!0.047$, which requires $b \lesssim 2.5$, may necessitate adequate 21-cm foreground mitigation techniques as proposed in recent literature~\citep[see e.g.][]{Gagnon-Hartman:2021erd}. 

In addition, we forecast for $\Lambda$CDM + $N_{\rm eff}$ + $Y_p$ model by varying both $N_{\rm eff}$ and $Y_p$ freely. We show the forecast results (68\% confidence level contours) for Foreground \Romannum{2} with regular feedback for 21-cm spectra in Fig.~\ref{fig:NeffYp}. By comparing blue and green contours, we conclude that additional information from 21-cm spectra indeed improves the sensitivity on all the cosmological parameters, in particular improving the error on $N_{\rm eff}$ by a factor of 1.91 {compared to combination of CMB-S4 and DESI BAO} in part by mitigating the strong degeneracy between $N_{\rm eff}$ and $Y_p$. The results of marginalised $1\sigma$ sensitivities to cosmological parameters are given in Table~\ref{tab:1sigma}, including those with Foreground \Romannum{1} setting as well as results for the $\Lambda$CDM + $N_{\rm eff}$ model. We further show the improvement factors $\left(\frac{\sigma_{\rm CMB+BAO}}{\sigma_{\rm CMB+BAO+21cm}}\right)$ in Table \ref{tab:improvements} for three different assumed CMB surveys, with regular feedback and Foreground \Romannum{2} setting.

\begin{figure}[!t]
\centering
\includegraphics[width=1\linewidth, trim= 20 10 20 10]{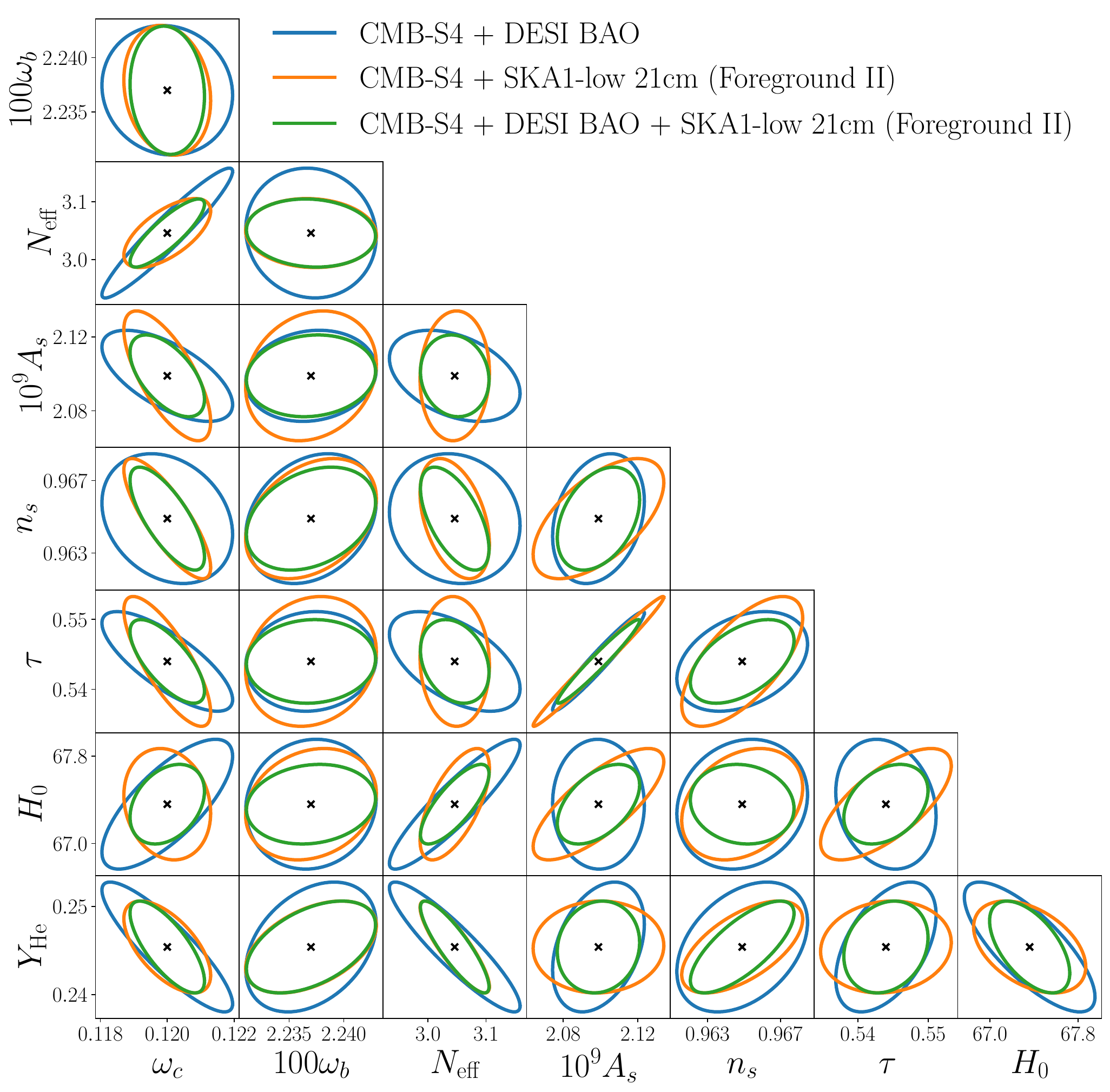}
\caption{Sensitivities to $\Lambda$CDM + $N_{\rm eff}$ + $Y_{\rm He}$. For 21-cm signals, Foreground \Romannum{2} and regular feedback are assumed.}
\label{fig:NeffYp}
\end{figure}
\begin{table}[!b]
  \centering
  \begin{tabular}{c||c|c|c|c}
     & \multicolumn{2}{c|}{Foreground \Romannum{1}} & \multicolumn{2}{c}{Foreground \Romannum{2}} \\
    \hline
    \hline
    $\sigma(\omega_c)$ & $5.41\times10^{-4}$ &$1.11\times10^{-3}$ & $4.72\times10^{-4}$ & $7.35\times10^{-4}$\\
    $\sigma(\omega_b)$ & $3.69\times 10^{-5}$ &$3.93\times 10^{-5}$ &$3.51\times 10^{-5}$ & $3.90\times 10^{-5}$\\
    $\sigma(N_{\rm eff})$ & 0.023 & 0.062 & 0.017 & 0.039\\
    $\sigma(A_s)$ & $1.50\times10^{-11}$ & $1.57\times10^{-11}$ & $1.45\times10^{-11}$ & $1.46\times10^{-11}$\\
    $\sigma(n_s)$ &$1.98\times10^{-3}$  & $2.26\times10^{-3}$ & $1.29\times10^{-3}$ & $1.88\times10^{-3}$\\
    $\sigma(\tau)$ &  $4.02\times10^{-3}$ & $4.47\times10^{-3}$ & $3.85\times10^{-3}$ & $3.95\times10^{-3}$\\
    $\sigma(H_0)$ & 0.22 & 0.34 & 0.19 & 0.24 \\
    \hline
    $\sigma(Y_{\rm He})$ & . & $4.40\times10^{-3}$ & . & $3.44\times10^{-3}$
  \end{tabular}
  \caption{Forecasts with CMB S4 + DESI BAO + 21cm SKA1-low for two different foregrounds, \{Foreground \Romannum{1}, Foreground \Romannum{2}\} corresponding to $(a,b) = (0.05\;\text{Mpc}^{-1}, 2)$ and $(0.01\;\text{Mpc}^{-1},1)$, which define $k_{\parallel,{\rm min}} = a + b\;k_\perp$. For each foreground, each column corresponds to sensitivities in $\Lambda$CDM + $N_{\rm eff}$ and $\Lambda$CDM + $N_{\rm eff}$ + $Y_{\rm He}$ models, respectively.}
  \label{tab:1sigma}
\end{table}
\begin{table}[hb!]
  \centering
  \begin{tabular}{c||c|c|c|c|c|c}
     & \multicolumn{2}{c|}{CMB-S4} & \multicolumn{2}{c|}{Advanced SO} & \multicolumn{2}{c}{SO} \\
    \hline
    \hline
    $\omega_c$  & ~~1.20~~ & ~~1.75~~ & ~~1.41~~ & ~~2.01~~ & ~~1.61~~ & ~~2.31~~\\
    $\omega_b$ & 1.07 & 1.01 & 1.10 & 1.02 & 1.15 & 1.03\\
    $N_{\rm eff}$ & 1.51 & 1.91 & 1.76 & 2.11 & 1.95 & 2.36\\
    $A_s$ & 1.04 & 1.11 & 1.05 & 1.13 & 1.05 & 1.14\\
    $n_s$   & 1.71 & 1.26 & 2.00 & 1.37 & 2.15 & 1.43\\
    $\tau$   & 1.05 & 1.20 & 1.07 & 1.24 & 1.07 & 1.27\\
    $H_0$   & 1.22 & 1.63 & 1.31 & 1.76 & 1.39 & 1.94\\
    \hline
    $Y_{\rm He}$ & . & 1.41 & . & 1.42 & . & 1.47
  \end{tabular}
  \caption{Improvement factors $\frac{\sigma_{\rm CMB+BAO}}{\sigma_{\rm CMB+BAO+21cm}}$ of CMB + DESI BAO + 21cm SKA1-low for three different CMB surveys with Foreground \Romannum{2} and regular feedback. For each foreground, each column corresponds to $\Lambda$CDM + $N_{\rm eff}$ and $\Lambda$CDM + $N_{\rm eff}$ + $Y_{\rm He}$ model, respectively.}
  \label{tab:improvements}
\end{table}
\begin{table}[hb!]
  \centering
  \begin{tabular}{c||c|c}
     & ~~~Foreground \Romannum{1}~~~ & ~~~Foreground \Romannum{2}~~~  \\
    \hline
    \hline
    High &\begin{tabular}{@{}c@{}} 0.023 \\ 0.062 (0.0044)\end{tabular} & \begin{tabular}{@{}c@{}} 0.017 \\ 0.042 (0.0037) \end{tabular}  \\
    \hline
    ~~Regular~~ & \begin{tabular}{@{}c@{}} 0.023 \\ 0.062 (0.0044)\end{tabular} & \begin{tabular}{@{}c@{}} 0.017 \\ 0.039 (0.0034) \end{tabular} \\
    \hline
    Low & \begin{tabular}{@{}c@{}} 0.023 \\ 0.059 (0.0042)\end{tabular} & \begin{tabular}{@{}c@{}} 0.016 \\ 0.035 (0.0031) \end{tabular} 
  \end{tabular}
  \caption{$1\sigma$ sensitivities of CMB S4 + DESI BAO + 21cm SKA1-low to $N_{\rm eff}$ with three feedback levels of \texttt{21cmvFAST} and two foreground settings defined in the main texts. Sensitivities to $Y_p$ are given in parentheses when varied freely, next to those to $N_{\rm eff}$.}
  \label{tab:feedback}
\end{table}

Since the level of the feedback remain largely uncertain during cosmic dawn, we consider a range of feedback levels that affect the modeling of the 21-cm power spectra from \texttt{21cmvFAST}. Our results (sensitivies on $N_{\rm eff}$ and $Y_p$) are summarized in Table~\ref{tab:feedback}. For each level of feedback, the first (second) rows indicate the marginalized $1\sigma$ sensitivity to $N_{\rm eff}$ ($N_{\rm eff}$, and $Y_p$ in parenthesis) for $\Lambda$CDM + $N_{\rm eff}$ ($\Lambda$CDM + $N_{\rm eff}$ + $Y_p$) model. We find the impact of different levels of feedback becomes more important in case of more optimistic foregrounds,
as the statistical power of the 21-cm experiment is greater. Nevertheless, we find the impact of the feedback on sensitivities to parameters is within $\lesssim10\%$ level overall.

\section{Discussion}

The velocity acoustic oscillation (VAO) signature in the 21-cm spectrum from cosmic dawn ($12\lesssim z\lesssim30$) provides a new window into probing the number of relativistic degrees of freedom $N_{\rm eff}$ (or equivalently, light relics from the early Universe) -- a key goal of upcoming Stage-4 cosmology experiments like CMB-S4\footnote{See \href{https://cmb-s4.org/science/the-dark-universe/}{cmb-s4.org/science/the-dark-universe}, for example.}. High-precision measurement of $N_{\rm eff}$ will grant a major advance in our understanding of the particle content and thermal history of our Universe.   

When taking advantage of upcoming 21-cm surveys, foregrounds removal will certainly play an important role. While we assume the usual `wedge' shape foreground avoidance, the methods for foreground subtraction and signal reconstruction have been proposed showing growing evidence that foregrounds mitigation can be improved in the future \cite{Beane:2018dzk,Li:2019znt,Villanueva-Domingo:2020wpt,Makinen:2020gvh,Gagnon-Hartman:2021erd}. We have also omitted taking into account correlations of 21-cm fluctuations between different redshift bins, which may impact the statistical power afforded by the largest scales, albeit possibly marginally. We did not model redshift-space distortion (RSD) due to peculiar velocities as well \cite{Bharadwaj:2004nr,Barkana:2004zy}, which we defer to the future work. However, given that the constraining power on $N_{\rm eff}$ is coming from the VAO signiture, we expect no significant impact on our results from RSD unless its effect is largely degenerate with the VAOs. In addition, our result is sensitive to survey volume as discussed in Ref.~\cite{Mellema:2012ht}, since the signal is cosmic variance dominated. In what follows, we could investigate these matters in more detail.  

Our ability to measure the VAO signature and cosmology from the cosmic dawn also depends on the astrophysics of this era, such as the LW feedback strength which remains largely uncharted. Here we assumed a range of phenomenological parameters [six fitting coefficients at each redshift defined in Eq.~\eqref{eq:fit} and \eqref{eq:poly}] to capture the effects of these complications before marginalizing over them in an information-matrix analysis in order to assess the cosmological information content of the 21-cm signal. In future analyses, we could extend the observations of this paper by conducting more robust simulations with comprehensive assumptions and modelling to further validate our results (see e.g., the code recently developed in Refs.~\cite{Munoz:2023kkg,Flitter:2023mjj,Flitter:2023rzv} as possible tools). Nevertheless, we anticipate an exciting future for the cosmological significance of the
cosmic-dawn signal, and its potential role in probing the light relics content of our Universe.\\

\section*{Acknowledgements}

We thank Yacine Ali-Ha\"imoud and Julian Mu\~{n}oz for useful comments on the draft. NL is supported by the Center for Cosmology and Particle Physics at New York University through the James Arthur Graduate Associate Fellowship. NL thanks Marc Kamionkowski, JHU Department of Physics, and Yacine Ali-Ha\"imoud for financial support during a visit to JHU. This work was supported in part through the NYU IT High Performance Computing resources, services, and staff expertise. This work was also carried out in part at the Advanced Research Computing at Hopkins (ARCH) core facility  (rockfish.jhu.edu), which is supported by the National Science Foundation (NSF) grant number OAC1920103. SCH is supported by the Horizon Fellowship from Johns Hopkins University. SCH thanks Julian Mu\~{n}oz and Jos\'e Luis Bernal for useful discussions. This work was performed in part at Aspen Center for Physics, which is supported by National Science Foundation grant PHY-2210452.

\bibliography{cosmicdawn}

\begin{thebibliography}{73}%
\makeatletter
\providecommand \@ifxundefined [1]{%
 \@ifx{#1\undefined}
}%
\providecommand \@ifnum [1]{%
 \ifnum #1\expandafter \@firstoftwo
 \else \expandafter \@secondoftwo
 \fi
}%
\providecommand \@ifx [1]{%
 \ifx #1\expandafter \@firstoftwo
 \else \expandafter \@secondoftwo
 \fi
}%
\providecommand \natexlab [1]{#1}%
\providecommand \enquote  [1]{``#1''}%
\providecommand \bibnamefont  [1]{#1}%
\providecommand \bibfnamefont [1]{#1}%
\providecommand \citenamefont [1]{#1}%
\providecommand \href@noop [0]{\@secondoftwo}%
\providecommand \href [0]{\begingroup \@sanitize@url \@href}%
\providecommand \@href[1]{\@@startlink{#1}\@@href}%
\providecommand \@@href[1]{\endgroup#1\@@endlink}%
\providecommand \@sanitize@url [0]{\catcode `\\12\catcode `\$12\catcode
  `\&12\catcode `\#12\catcode `\^12\catcode `\_12\catcode `\%12\relax}%
\providecommand \@@startlink[1]{}%
\providecommand \@@endlink[0]{}%
\providecommand \url  [0]{\begingroup\@sanitize@url \@url }%
\providecommand \@url [1]{\endgroup\@href {#1}{\urlprefix }}%
\providecommand \urlprefix  [0]{URL }%
\providecommand \Eprint [0]{\href }%
\providecommand \doibase [0]{http://dx.doi.org/}%
\providecommand \selectlanguage [0]{\@gobble}%
\providecommand \bibinfo  [0]{\@secondoftwo}%
\providecommand \bibfield  [0]{\@secondoftwo}%
\providecommand \translation [1]{[#1]}%
\providecommand \BibitemOpen [0]{}%
\providecommand \bibitemStop [0]{}%
\providecommand \bibitemNoStop [0]{.\EOS\space}%
\providecommand \EOS [0]{\spacefactor3000\relax}%
\providecommand \BibitemShut  [1]{\csname bibitem#1\endcsname}%
\let\auto@bib@innerbib\@empty
\bibitem [{\citenamefont {Abazajian}\ \emph {et~al.}(2016)\citenamefont
  {Abazajian} \emph {et~al.}}]{CMB-S4:2016ple}%
  \BibitemOpen
  \bibfield  {author} {\bibinfo {author} {\bibfnamefont {K.~N.}\ \bibnamefont
  {Abazajian}} \emph {et~al.} (\bibinfo {collaboration} {CMB-S4}),\ }\href@noop
  {} {\  (\bibinfo {year} {2016})},\ \Eprint {http://arxiv.org/abs/1610.02743}
  {arXiv:1610.02743 [astro-ph.CO]} \BibitemShut {NoStop}%
\bibitem [{\citenamefont {Abazajian}\ \emph {et~al.}(2019)\citenamefont
  {Abazajian} \emph {et~al.}}]{Abazajian:2019eic}%
  \BibitemOpen
  \bibfield  {author} {\bibinfo {author} {\bibfnamefont {K.}~\bibnamefont
  {Abazajian}} \emph {et~al.},\ }\href@noop {} {\  (\bibinfo {year} {2019})},\
  \Eprint {http://arxiv.org/abs/1907.04473} {arXiv:1907.04473 [astro-ph.IM]}
  \BibitemShut {NoStop}%
\bibitem [{\citenamefont {Abazajian}\ \emph {et~al.}(2022)\citenamefont
  {Abazajian} \emph {et~al.}}]{Abazajian:2022nyh}%
  \BibitemOpen
  \bibfield  {author} {\bibinfo {author} {\bibfnamefont {K.}~\bibnamefont
  {Abazajian}} \emph {et~al.},\ }in\ \href@noop {} {\emph {\bibinfo {booktitle}
  {{2022 Snowmass Summer Study}}}}\ (\bibinfo {year} {2022})\ \Eprint
  {http://arxiv.org/abs/2203.08024} {arXiv:2203.08024 [astro-ph.CO]}
  \BibitemShut {NoStop}%
\bibitem [{\citenamefont {Aguirre}\ \emph {et~al.}(2018)\citenamefont {Aguirre}
  \emph {et~al.}}]{Ade:2018sbj}%
  \BibitemOpen
  \bibfield  {author} {\bibinfo {author} {\bibfnamefont {J.}~\bibnamefont
  {Aguirre}} \emph {et~al.} (\bibinfo {collaboration} {Simons Observatory}),\
  }\href@noop {} {\  (\bibinfo {year} {2018})},\ \Eprint
  {http://arxiv.org/abs/1808.07445} {arXiv:1808.07445 [astro-ph.CO]}
  \BibitemShut {NoStop}%
\bibitem [{\citenamefont {Abitbol}\ \emph {et~al.}(2019)\citenamefont {Abitbol}
  \emph {et~al.}}]{SimonsObservatory:2019qwx}%
  \BibitemOpen
  \bibfield  {author} {\bibinfo {author} {\bibfnamefont {M.~H.}\ \bibnamefont
  {Abitbol}} \emph {et~al.} (\bibinfo {collaboration} {Simons Observatory}),\
  }\href@noop {} {\bibfield  {journal} {\bibinfo  {journal} {Bull. Am. Astron.
  Soc.}\ }\textbf {\bibinfo {volume} {51}},\ \bibinfo {pages} {147} (\bibinfo
  {year} {2019})},\ \Eprint {http://arxiv.org/abs/1907.08284} {arXiv:1907.08284
  [astro-ph.IM]} \BibitemShut {NoStop}%
\bibitem [{\citenamefont {Sehgal}\ \emph {et~al.}(2019)\citenamefont {Sehgal}
  \emph {et~al.}}]{Sehgal:2019ewc}%
  \BibitemOpen
  \bibfield  {author} {\bibinfo {author} {\bibfnamefont {N.}~\bibnamefont
  {Sehgal}} \emph {et~al.},\ }\href@noop {} {\  (\bibinfo {year} {2019})},\
  \Eprint {http://arxiv.org/abs/1906.10134} {arXiv:1906.10134 [astro-ph.CO]}
  \BibitemShut {NoStop}%
\bibitem [{\citenamefont {Aiola}\ \emph {et~al.}(2022)\citenamefont {Aiola}
  \emph {et~al.}}]{CMB-HD:2022bsz}%
  \BibitemOpen
  \bibfield  {author} {\bibinfo {author} {\bibfnamefont {S.}~\bibnamefont
  {Aiola}} \emph {et~al.} (\bibinfo {collaboration} {CMB-HD}),\ }\href@noop {}
  {\  (\bibinfo {year} {2022})},\ \Eprint {http://arxiv.org/abs/2203.05728}
  {arXiv:2203.05728 [astro-ph.CO]} \BibitemShut {NoStop}%
\bibitem [{\citenamefont {Aghamousa}\ \emph {et~al.}(2016)\citenamefont
  {Aghamousa} \emph {et~al.}}]{DESI:2016fyo}%
  \BibitemOpen
  \bibfield  {author} {\bibinfo {author} {\bibfnamefont {A.}~\bibnamefont
  {Aghamousa}} \emph {et~al.} (\bibinfo {collaboration} {DESI}),\ }\href@noop
  {} {\  (\bibinfo {year} {2016})},\ \Eprint {http://arxiv.org/abs/1611.00036}
  {arXiv:1611.00036 [astro-ph.IM]} \BibitemShut {NoStop}%
\bibitem [{\citenamefont {Levi}\ \emph {et~al.}(2019)\citenamefont {Levi} \emph
  {et~al.}}]{DESI:2019jxc}%
  \BibitemOpen
  \bibfield  {author} {\bibinfo {author} {\bibfnamefont {M.~E.}\ \bibnamefont
  {Levi}} \emph {et~al.} (\bibinfo {collaboration} {DESI}),\ }\href@noop {} {\
  (\bibinfo {year} {2019})},\ \Eprint {http://arxiv.org/abs/1907.10688}
  {arXiv:1907.10688 [astro-ph.IM]} \BibitemShut {NoStop}%
\bibitem [{\citenamefont {Abate}\ \emph {et~al.}(2012)\citenamefont {Abate}
  \emph {et~al.}}]{LSSTDarkEnergyScience:2012kar}%
  \BibitemOpen
  \bibfield  {author} {\bibinfo {author} {\bibfnamefont {A.}~\bibnamefont
  {Abate}} \emph {et~al.} (\bibinfo {collaboration} {LSST Dark Energy
  Science}),\ }\href@noop {} {\  (\bibinfo {year} {2012})},\ \Eprint
  {http://arxiv.org/abs/1211.0310} {arXiv:1211.0310 [astro-ph.CO]} \BibitemShut
  {NoStop}%
\bibitem [{\citenamefont {{The LSST Dark Energy Science Collaboration}}\ \emph
  {et~al.}(2018)\citenamefont {{The LSST Dark Energy Science Collaboration}},
  \citenamefont {{Mandelbaum}}, \citenamefont {{Eifler}}, \citenamefont
  {{Hlo{\v{z}}ek}}, \citenamefont {{Collett}}, \citenamefont {{Gawiser}},\ and\
  \citenamefont {{Scolnic}}}]{LSSTDarkEnergyScience:2018jkl}%
  \BibitemOpen
  \bibfield  {author} {\bibinfo {author} {\bibnamefont {{The LSST Dark Energy
  Science Collaboration}}}, \bibinfo {author} {\bibfnamefont {R.}~\bibnamefont
  {{Mandelbaum}}}, \bibinfo {author} {\bibfnamefont {T.}~\bibnamefont
  {{Eifler}}}, \bibinfo {author} {\bibfnamefont {R.}~\bibnamefont
  {{Hlo{\v{z}}ek}}}, \bibinfo {author} {\bibfnamefont {T.}~\bibnamefont
  {{Collett}}}, \bibinfo {author} {\bibfnamefont {E.}~\bibnamefont
  {{Gawiser}}}, \ and\ \bibinfo {author} {\bibfnamefont {o.}~\bibnamefont
  {{Scolnic}}},\ }\href@noop {} {\bibfield  {journal} {\bibinfo  {journal}
  {arXiv e-prints}\ ,\ \bibinfo {eid} {arXiv:1809.01669}} (\bibinfo {year}
  {2018})},\ \Eprint {http://arxiv.org/abs/1809.01669} {arXiv:1809.01669
  [astro-ph.CO]} \BibitemShut {NoStop}%
\bibitem [{\citenamefont {Ivezi\'c}\ \emph {et~al.}(2019)\citenamefont
  {Ivezi\'c} \emph {et~al.}}]{LSST:2008ijt}%
  \BibitemOpen
  \bibfield  {author} {\bibinfo {author} {\bibfnamefont {v.}~\bibnamefont
  {Ivezi\'c}} \emph {et~al.} (\bibinfo {collaboration} {LSST}),\ }\href
  {\doibase 10.3847/1538-4357/ab042c} {\bibfield  {journal} {\bibinfo
  {journal} {Astrophys. J.}\ }\textbf {\bibinfo {volume} {873}},\ \bibinfo
  {pages} {111} (\bibinfo {year} {2019})},\ \Eprint
  {http://arxiv.org/abs/0805.2366} {arXiv:0805.2366 [astro-ph]} \BibitemShut
  {NoStop}%
\bibitem [{\citenamefont {Spergel}\ \emph {et~al.}(2015)\citenamefont {Spergel}
  \emph {et~al.}}]{Spergel:2015sza}%
  \BibitemOpen
  \bibfield  {author} {\bibinfo {author} {\bibfnamefont {D.}~\bibnamefont
  {Spergel}} \emph {et~al.},\ }\href@noop {} {\  (\bibinfo {year} {2015})},\
  \Eprint {http://arxiv.org/abs/1503.03757} {arXiv:1503.03757 [astro-ph.IM]}
  \BibitemShut {NoStop}%
\bibitem [{\citenamefont {Laureijs}\ \emph {et~al.}(2011)\citenamefont
  {Laureijs} \emph {et~al.}}]{EUCLID:2011zbd}%
  \BibitemOpen
  \bibfield  {author} {\bibinfo {author} {\bibfnamefont {R.}~\bibnamefont
  {Laureijs}} \emph {et~al.} (\bibinfo {collaboration} {EUCLID}),\ }\href@noop
  {} {\  (\bibinfo {year} {2011})},\ \Eprint {http://arxiv.org/abs/1110.3193}
  {arXiv:1110.3193 [astro-ph.CO]} \BibitemShut {NoStop}%
\bibitem [{\citenamefont {Bashinsky}\ and\ \citenamefont
  {Seljak}(2004)}]{Bashinsky:2003tk}%
  \BibitemOpen
  \bibfield  {author} {\bibinfo {author} {\bibfnamefont {S.}~\bibnamefont
  {Bashinsky}}\ and\ \bibinfo {author} {\bibfnamefont {U.}~\bibnamefont
  {Seljak}},\ }\href {\doibase 10.1103/PhysRevD.69.083002} {\bibfield
  {journal} {\bibinfo  {journal} {Phys. Rev. D}\ }\textbf {\bibinfo {volume}
  {69}},\ \bibinfo {pages} {083002} (\bibinfo {year} {2004})},\ \Eprint
  {http://arxiv.org/abs/astro-ph/0310198} {arXiv:astro-ph/0310198} \BibitemShut
  {NoStop}%
\bibitem [{\citenamefont {Hou}\ \emph {et~al.}(2013)\citenamefont {Hou},
  \citenamefont {Keisler}, \citenamefont {Knox}, \citenamefont {Millea},\ and\
  \citenamefont {Reichardt}}]{Hou:2011ec}%
  \BibitemOpen
  \bibfield  {author} {\bibinfo {author} {\bibfnamefont {Z.}~\bibnamefont
  {Hou}}, \bibinfo {author} {\bibfnamefont {R.}~\bibnamefont {Keisler}},
  \bibinfo {author} {\bibfnamefont {L.}~\bibnamefont {Knox}}, \bibinfo {author}
  {\bibfnamefont {M.}~\bibnamefont {Millea}}, \ and\ \bibinfo {author}
  {\bibfnamefont {C.}~\bibnamefont {Reichardt}},\ }\href {\doibase
  10.1103/PhysRevD.87.083008} {\bibfield  {journal} {\bibinfo  {journal} {Phys.
  Rev. D}\ }\textbf {\bibinfo {volume} {87}},\ \bibinfo {pages} {083008}
  (\bibinfo {year} {2013})},\ \Eprint {http://arxiv.org/abs/1104.2333}
  {arXiv:1104.2333 [astro-ph.CO]} \BibitemShut {NoStop}%
\bibitem [{\citenamefont {Follin}\ \emph {et~al.}(2015)\citenamefont {Follin},
  \citenamefont {Knox}, \citenamefont {Millea},\ and\ \citenamefont
  {Pan}}]{Follin:2015hya}%
  \BibitemOpen
  \bibfield  {author} {\bibinfo {author} {\bibfnamefont {B.}~\bibnamefont
  {Follin}}, \bibinfo {author} {\bibfnamefont {L.}~\bibnamefont {Knox}},
  \bibinfo {author} {\bibfnamefont {M.}~\bibnamefont {Millea}}, \ and\ \bibinfo
  {author} {\bibfnamefont {Z.}~\bibnamefont {Pan}},\ }\href {\doibase
  10.1103/PhysRevLett.115.091301} {\bibfield  {journal} {\bibinfo  {journal}
  {Phys. Rev. Lett.}\ }\textbf {\bibinfo {volume} {115}},\ \bibinfo {pages}
  {091301} (\bibinfo {year} {2015})},\ \Eprint
  {http://arxiv.org/abs/1503.07863} {arXiv:1503.07863 [astro-ph.CO]}
  \BibitemShut {NoStop}%
\bibitem [{\citenamefont {Bashinsky}\ and\ \citenamefont
  {Bertschinger}(2002)}]{Bashinsky:2002vx}%
  \BibitemOpen
  \bibfield  {author} {\bibinfo {author} {\bibfnamefont {S.}~\bibnamefont
  {Bashinsky}}\ and\ \bibinfo {author} {\bibfnamefont {E.}~\bibnamefont
  {Bertschinger}},\ }\href {\doibase 10.1103/PhysRevD.65.123008} {\bibfield
  {journal} {\bibinfo  {journal} {Phys. Rev. D}\ }\textbf {\bibinfo {volume}
  {65}},\ \bibinfo {pages} {123008} (\bibinfo {year} {2002})},\ \Eprint
  {http://arxiv.org/abs/astro-ph/0202215} {arXiv:astro-ph/0202215} \BibitemShut
  {NoStop}%
\bibitem [{\citenamefont {Eisenstein}\ \emph {et~al.}(2007)\citenamefont
  {Eisenstein}, \citenamefont {Seo},\ and\ \citenamefont
  {White}}]{Eisenstein:2006nj}%
  \BibitemOpen
  \bibfield  {author} {\bibinfo {author} {\bibfnamefont {D.~J.}\ \bibnamefont
  {Eisenstein}}, \bibinfo {author} {\bibfnamefont {H.-j.}\ \bibnamefont {Seo}},
  \ and\ \bibinfo {author} {\bibfnamefont {M.~J.}\ \bibnamefont {White}},\
  }\href {\doibase 10.1086/518755} {\bibfield  {journal} {\bibinfo  {journal}
  {Astrophys. J.}\ }\textbf {\bibinfo {volume} {664}},\ \bibinfo {pages} {660}
  (\bibinfo {year} {2007})},\ \Eprint {http://arxiv.org/abs/astro-ph/0604361}
  {arXiv:astro-ph/0604361} \BibitemShut {NoStop}%
\bibitem [{\citenamefont {Baumann}\ \emph {et~al.}(2017)\citenamefont
  {Baumann}, \citenamefont {Green},\ and\ \citenamefont
  {Zaldarriaga}}]{Baumann:2017lmt}%
  \BibitemOpen
  \bibfield  {author} {\bibinfo {author} {\bibfnamefont {D.}~\bibnamefont
  {Baumann}}, \bibinfo {author} {\bibfnamefont {D.}~\bibnamefont {Green}}, \
  and\ \bibinfo {author} {\bibfnamefont {M.}~\bibnamefont {Zaldarriaga}},\
  }\href {\doibase 10.1088/1475-7516/2017/11/007} {\bibfield  {journal}
  {\bibinfo  {journal} {JCAP}\ }\textbf {\bibinfo {volume} {11}},\ \bibinfo
  {pages} {007} (\bibinfo {year} {2017})},\ \Eprint
  {http://arxiv.org/abs/1703.00894} {arXiv:1703.00894 [astro-ph.CO]}
  \BibitemShut {NoStop}%
\bibitem [{\citenamefont {Baumann}\ \emph {et~al.}(2018)\citenamefont
  {Baumann}, \citenamefont {Green},\ and\ \citenamefont
  {Wallisch}}]{Baumann:2017gkg}%
  \BibitemOpen
  \bibfield  {author} {\bibinfo {author} {\bibfnamefont {D.}~\bibnamefont
  {Baumann}}, \bibinfo {author} {\bibfnamefont {D.}~\bibnamefont {Green}}, \
  and\ \bibinfo {author} {\bibfnamefont {B.}~\bibnamefont {Wallisch}},\ }\href
  {\doibase 10.1088/1475-7516/2018/08/029} {\bibfield  {journal} {\bibinfo
  {journal} {JCAP}\ }\textbf {\bibinfo {volume} {08}},\ \bibinfo {pages} {029}
  (\bibinfo {year} {2018})},\ \Eprint {http://arxiv.org/abs/1712.08067}
  {arXiv:1712.08067 [astro-ph.CO]} \BibitemShut {NoStop}%
\bibitem [{\citenamefont {Baumann}\ \emph {et~al.}(2019)\citenamefont
  {Baumann}, \citenamefont {Beutler}, \citenamefont {Flauger}, \citenamefont
  {Green}, \citenamefont {Slosar}, \citenamefont {Vargas-Maga\~na},
  \citenamefont {Wallisch},\ and\ \citenamefont {Y\`eche}}]{Baumann:2019keh}%
  \BibitemOpen
  \bibfield  {author} {\bibinfo {author} {\bibfnamefont {D.}~\bibnamefont
  {Baumann}}, \bibinfo {author} {\bibfnamefont {F.}~\bibnamefont {Beutler}},
  \bibinfo {author} {\bibfnamefont {R.}~\bibnamefont {Flauger}}, \bibinfo
  {author} {\bibfnamefont {D.}~\bibnamefont {Green}}, \bibinfo {author}
  {\bibfnamefont {A.}~\bibnamefont {Slosar}}, \bibinfo {author} {\bibfnamefont
  {M.}~\bibnamefont {Vargas-Maga\~na}}, \bibinfo {author} {\bibfnamefont
  {B.}~\bibnamefont {Wallisch}}, \ and\ \bibinfo {author} {\bibfnamefont
  {C.}~\bibnamefont {Y\`eche}},\ }\href {\doibase 10.1038/s41567-019-0435-6}
  {\bibfield  {journal} {\bibinfo  {journal} {Nature Phys.}\ }\textbf {\bibinfo
  {volume} {15}},\ \bibinfo {pages} {465} (\bibinfo {year} {2019})},\ \Eprint
  {http://arxiv.org/abs/1803.10741} {arXiv:1803.10741 [astro-ph.CO]}
  \BibitemShut {NoStop}%
\bibitem [{\citenamefont {Tseliakhovich}\ and\ \citenamefont
  {Hirata}(2010)}]{Tseliakhovich:2010bj}%
  \BibitemOpen
  \bibfield  {author} {\bibinfo {author} {\bibfnamefont {D.}~\bibnamefont
  {Tseliakhovich}}\ and\ \bibinfo {author} {\bibfnamefont {C.}~\bibnamefont
  {Hirata}},\ }\href {\doibase 10.1103/PhysRevD.82.083520} {\bibfield
  {journal} {\bibinfo  {journal} {Phys. Rev. D}\ }\textbf {\bibinfo {volume}
  {82}},\ \bibinfo {pages} {083520} (\bibinfo {year} {2010})},\ \Eprint
  {http://arxiv.org/abs/1005.2416} {arXiv:1005.2416 [astro-ph.CO]} \BibitemShut
  {NoStop}%
\bibitem [{\citenamefont {Muñoz}(2019{\natexlab{a}})}]{Munoz:2019rhi}%
  \BibitemOpen
  \bibfield  {author} {\bibinfo {author} {\bibfnamefont {J.~B.}\ \bibnamefont
  {Muñoz}},\ }\href {\doibase 10.1103/PhysRevD.100.063538} {\bibfield
  {journal} {\bibinfo  {journal} {Phys. Rev.}\ }\textbf {\bibinfo {volume}
  {D100}},\ \bibinfo {pages} {063538} (\bibinfo {year} {2019}{\natexlab{a}})},\
  \Eprint {http://arxiv.org/abs/1904.07881} {arXiv:1904.07881 [astro-ph.CO]}
  \BibitemShut {NoStop}%
\bibitem [{\citenamefont {Machacek}\ \emph {et~al.}(2001)\citenamefont
  {Machacek}, \citenamefont {Bryan},\ and\ \citenamefont
  {Abel}}]{Machacek:2000us}%
  \BibitemOpen
  \bibfield  {author} {\bibinfo {author} {\bibfnamefont {M.~E.}\ \bibnamefont
  {Machacek}}, \bibinfo {author} {\bibfnamefont {G.~L.}\ \bibnamefont {Bryan}},
  \ and\ \bibinfo {author} {\bibfnamefont {T.}~\bibnamefont {Abel}},\ }\href
  {\doibase 10.1086/319014} {\bibfield  {journal} {\bibinfo  {journal}
  {Astrophys. J.}\ }\textbf {\bibinfo {volume} {548}},\ \bibinfo {pages} {509}
  (\bibinfo {year} {2001})},\ \Eprint {http://arxiv.org/abs/astro-ph/0007198}
  {arXiv:astro-ph/0007198} \BibitemShut {NoStop}%
\bibitem [{\citenamefont {Fialkov}\ \emph {et~al.}(2013)\citenamefont
  {Fialkov}, \citenamefont {Barkana}, \citenamefont {Visbal}, \citenamefont
  {Tseliakhovich},\ and\ \citenamefont {Hirata}}]{Fialkov:2012su}%
  \BibitemOpen
  \bibfield  {author} {\bibinfo {author} {\bibfnamefont {A.}~\bibnamefont
  {Fialkov}}, \bibinfo {author} {\bibfnamefont {R.}~\bibnamefont {Barkana}},
  \bibinfo {author} {\bibfnamefont {E.}~\bibnamefont {Visbal}}, \bibinfo
  {author} {\bibfnamefont {D.}~\bibnamefont {Tseliakhovich}}, \ and\ \bibinfo
  {author} {\bibfnamefont {C.~M.}\ \bibnamefont {Hirata}},\ }\href {\doibase
  10.1093/mnras/stt650} {\bibfield  {journal} {\bibinfo  {journal} {Mon. Not.
  Roy. Astron. Soc.}\ }\textbf {\bibinfo {volume} {432}},\ \bibinfo {pages}
  {2909} (\bibinfo {year} {2013})},\ \Eprint {http://arxiv.org/abs/1212.0513}
  {arXiv:1212.0513 [astro-ph.CO]} \BibitemShut {NoStop}%
\bibitem [{\citenamefont {Visbal}\ \emph {et~al.}(2014)\citenamefont {Visbal},
  \citenamefont {Haiman}, \citenamefont {Terrazas}, \citenamefont {Bryan},\
  and\ \citenamefont {Barkana}}]{Visbal:2014fta}%
  \BibitemOpen
  \bibfield  {author} {\bibinfo {author} {\bibfnamefont {E.}~\bibnamefont
  {Visbal}}, \bibinfo {author} {\bibfnamefont {Z.}~\bibnamefont {Haiman}},
  \bibinfo {author} {\bibfnamefont {B.}~\bibnamefont {Terrazas}}, \bibinfo
  {author} {\bibfnamefont {G.~L.}\ \bibnamefont {Bryan}}, \ and\ \bibinfo
  {author} {\bibfnamefont {R.}~\bibnamefont {Barkana}},\ }\href {\doibase
  10.1093/mnras/stu1710} {\bibfield  {journal} {\bibinfo  {journal} {Mon. Not.
  Roy. Astron. Soc.}\ }\textbf {\bibinfo {volume} {445}},\ \bibinfo {pages}
  {107} (\bibinfo {year} {2014})},\ \Eprint {http://arxiv.org/abs/1402.0882}
  {arXiv:1402.0882 [astro-ph.CO]} \BibitemShut {NoStop}%
\bibitem [{\citenamefont {Muñoz}(2019{\natexlab{b}})}]{Munoz:2019fkt}%
  \BibitemOpen
  \bibfield  {author} {\bibinfo {author} {\bibfnamefont {J.~B.}\ \bibnamefont
  {Muñoz}},\ }\href {\doibase 10.1103/PhysRevLett.123.131301} {\bibfield
  {journal} {\bibinfo  {journal} {Phys. Rev. Lett.}\ }\textbf {\bibinfo
  {volume} {123}},\ \bibinfo {pages} {131301} (\bibinfo {year}
  {2019}{\natexlab{b}})},\ \Eprint {http://arxiv.org/abs/1904.07868}
  {arXiv:1904.07868 [astro-ph.CO]} \BibitemShut {NoStop}%
\bibitem [{\citenamefont {Hotinli}\ \emph
  {et~al.}(2022{\natexlab{a}})\citenamefont {Hotinli}, \citenamefont {Marsh},\
  and\ \citenamefont {Kamionkowski}}]{Hotinli:2021vxg}%
  \BibitemOpen
  \bibfield  {author} {\bibinfo {author} {\bibfnamefont {S.~C.}\ \bibnamefont
  {Hotinli}}, \bibinfo {author} {\bibfnamefont {D.~J.~E.}\ \bibnamefont
  {Marsh}}, \ and\ \bibinfo {author} {\bibfnamefont {M.}~\bibnamefont
  {Kamionkowski}},\ }\href {\doibase 10.1103/PhysRevD.106.043529} {\bibfield
  {journal} {\bibinfo  {journal} {Phys. Rev. D}\ }\textbf {\bibinfo {volume}
  {106}},\ \bibinfo {pages} {043529} (\bibinfo {year} {2022}{\natexlab{a}})},\
  \Eprint {http://arxiv.org/abs/2112.06943} {arXiv:2112.06943 [astro-ph.CO]}
  \BibitemShut {NoStop}%
\bibitem [{\citenamefont {Hotinli}\ \emph
  {et~al.}(2021{\natexlab{a}})\citenamefont {Hotinli}, \citenamefont {Binnie},
  \citenamefont {Mu\~noz}, \citenamefont {Dinda},\ and\ \citenamefont
  {Kamionkowski}}]{Hotinli:2021xln}%
  \BibitemOpen
  \bibfield  {author} {\bibinfo {author} {\bibfnamefont {S.~C.}\ \bibnamefont
  {Hotinli}}, \bibinfo {author} {\bibfnamefont {T.}~\bibnamefont {Binnie}},
  \bibinfo {author} {\bibfnamefont {J.~B.}\ \bibnamefont {Mu\~noz}}, \bibinfo
  {author} {\bibfnamefont {B.~R.}\ \bibnamefont {Dinda}}, \ and\ \bibinfo
  {author} {\bibfnamefont {M.}~\bibnamefont {Kamionkowski}},\ }\href {\doibase
  10.1103/PhysRevD.104.063536} {\bibfield  {journal} {\bibinfo  {journal}
  {Phys. Rev. D}\ }\textbf {\bibinfo {volume} {104}},\ \bibinfo {pages}
  {063536} (\bibinfo {year} {2021}{\natexlab{a}})},\ \Eprint
  {http://arxiv.org/abs/2106.11979} {arXiv:2106.11979 [astro-ph.CO]}
  \BibitemShut {NoStop}%
\bibitem [{\citenamefont {Hotinli}\ \emph {et~al.}(2023)\citenamefont
  {Hotinli}, \citenamefont {Ferraro}, \citenamefont {Holder}, \citenamefont
  {Johnson}, \citenamefont {Kamionkowski},\ and\ \citenamefont
  {La~Plante}}]{Hotinli:2022jna}%
  \BibitemOpen
  \bibfield  {author} {\bibinfo {author} {\bibfnamefont {S.~C.}\ \bibnamefont
  {Hotinli}}, \bibinfo {author} {\bibfnamefont {S.}~\bibnamefont {Ferraro}},
  \bibinfo {author} {\bibfnamefont {G.~P.}\ \bibnamefont {Holder}}, \bibinfo
  {author} {\bibfnamefont {M.~C.}\ \bibnamefont {Johnson}}, \bibinfo {author}
  {\bibfnamefont {M.}~\bibnamefont {Kamionkowski}}, \ and\ \bibinfo {author}
  {\bibfnamefont {P.}~\bibnamefont {La~Plante}},\ }\href {\doibase
  10.1103/PhysRevD.107.103517} {\bibfield  {journal} {\bibinfo  {journal}
  {Phys. Rev. D}\ }\textbf {\bibinfo {volume} {107}},\ \bibinfo {pages}
  {103517} (\bibinfo {year} {2023})},\ \Eprint
  {http://arxiv.org/abs/2207.07660} {arXiv:2207.07660 [astro-ph.CO]}
  \BibitemShut {NoStop}%
\bibitem [{\citenamefont {Hotinli}(2023)}]{Hotinli:2022jnt}%
  \BibitemOpen
  \bibfield  {author} {\bibinfo {author} {\bibfnamefont {S.~C.}\ \bibnamefont
  {Hotinli}},\ }\href {\doibase 10.1103/PhysRevD.108.043528} {\bibfield
  {journal} {\bibinfo  {journal} {Phys. Rev. D}\ }\textbf {\bibinfo {volume}
  {108}},\ \bibinfo {pages} {043528} (\bibinfo {year} {2023})},\ \Eprint
  {http://arxiv.org/abs/2212.08004} {arXiv:2212.08004 [astro-ph.CO]}
  \BibitemShut {NoStop}%
\bibitem [{\citenamefont {Bowman}\ \emph {et~al.}(2018)\citenamefont {Bowman},
  \citenamefont {Rogers}, \citenamefont {Monsalve}, \citenamefont {Mozdzen},\
  and\ \citenamefont {Mahesh}}]{Bowman:2018yin}%
  \BibitemOpen
  \bibfield  {author} {\bibinfo {author} {\bibfnamefont {J.~D.}\ \bibnamefont
  {Bowman}}, \bibinfo {author} {\bibfnamefont {A.~E.~E.}\ \bibnamefont
  {Rogers}}, \bibinfo {author} {\bibfnamefont {R.~A.}\ \bibnamefont
  {Monsalve}}, \bibinfo {author} {\bibfnamefont {T.~J.}\ \bibnamefont
  {Mozdzen}}, \ and\ \bibinfo {author} {\bibfnamefont {N.}~\bibnamefont
  {Mahesh}},\ }\href {\doibase 10.1038/nature25792} {\bibfield  {journal}
  {\bibinfo  {journal} {Nature}\ }\textbf {\bibinfo {volume} {555}},\ \bibinfo
  {pages} {67} (\bibinfo {year} {2018})},\ \Eprint
  {http://arxiv.org/abs/1810.05912} {arXiv:1810.05912 [astro-ph.CO]}
  \BibitemShut {NoStop}%
\bibitem [{\citenamefont {Singh}\ \emph {et~al.}(2018)\citenamefont {Singh},
  \citenamefont {Subrahmanyan}, \citenamefont {Shankar}, \citenamefont {Rao},
  \citenamefont {Girish}, \citenamefont {Raghunathan}, \citenamefont
  {Somashekar},\ and\ \citenamefont {Srivani}}]{Singh:2017syr}%
  \BibitemOpen
  \bibfield  {author} {\bibinfo {author} {\bibfnamefont {S.}~\bibnamefont
  {Singh}}, \bibinfo {author} {\bibfnamefont {R.}~\bibnamefont {Subrahmanyan}},
  \bibinfo {author} {\bibfnamefont {N.~U.}\ \bibnamefont {Shankar}}, \bibinfo
  {author} {\bibfnamefont {M.~S.}\ \bibnamefont {Rao}}, \bibinfo {author}
  {\bibfnamefont {B.~S.}\ \bibnamefont {Girish}}, \bibinfo {author}
  {\bibfnamefont {A.}~\bibnamefont {Raghunathan}}, \bibinfo {author}
  {\bibfnamefont {R.}~\bibnamefont {Somashekar}}, \ and\ \bibinfo {author}
  {\bibfnamefont {K.~S.}\ \bibnamefont {Srivani}},\ }\href {\doibase
  10.1007/s10686-018-9584-3} {\bibfield  {journal} {\bibinfo  {journal} {Exper.
  Astron.}\ }\textbf {\bibinfo {volume} {45}},\ \bibinfo {pages} {269}
  (\bibinfo {year} {2018})},\ \Eprint {http://arxiv.org/abs/1710.01101}
  {arXiv:1710.01101 [astro-ph.IM]} \BibitemShut {NoStop}%
\bibitem [{\citenamefont {{Philip}}\ \emph {et~al.}(2019)\citenamefont
  {{Philip}}, \citenamefont {{Abdurashidova}}, \citenamefont {{Chiang}},
  \citenamefont {{Ghazi}}, \citenamefont {{Gumba}}, \citenamefont
  {{Heilgendorff}}, \citenamefont {{J{\'a}uregui-Garc{\'\i}a}}, \citenamefont
  {{Malepe}}, \citenamefont {{Nunhokee}}, \citenamefont {{Peterson}},
  \citenamefont {{Sievers}}, \citenamefont {{Simes}},\ and\ \citenamefont
  {{Spann}}}]{2019JAI.....850004P}%
  \BibitemOpen
  \bibfield  {author} {\bibinfo {author} {\bibfnamefont {L.}~\bibnamefont
  {{Philip}}}, \bibinfo {author} {\bibfnamefont {Z.}~\bibnamefont
  {{Abdurashidova}}}, \bibinfo {author} {\bibfnamefont {H.~C.}\ \bibnamefont
  {{Chiang}}}, \bibinfo {author} {\bibfnamefont {N.}~\bibnamefont {{Ghazi}}},
  \bibinfo {author} {\bibfnamefont {A.}~\bibnamefont {{Gumba}}}, \bibinfo
  {author} {\bibfnamefont {H.~M.}\ \bibnamefont {{Heilgendorff}}}, \bibinfo
  {author} {\bibfnamefont {J.~M.}\ \bibnamefont {{J{\'a}uregui-Garc{\'\i}a}}},
  \bibinfo {author} {\bibfnamefont {K.}~\bibnamefont {{Malepe}}}, \bibinfo
  {author} {\bibfnamefont {C.~D.}\ \bibnamefont {{Nunhokee}}}, \bibinfo
  {author} {\bibfnamefont {J.}~\bibnamefont {{Peterson}}}, \bibinfo {author}
  {\bibfnamefont {J.~L.}\ \bibnamefont {{Sievers}}}, \bibinfo {author}
  {\bibfnamefont {V.}~\bibnamefont {{Simes}}}, \ and\ \bibinfo {author}
  {\bibfnamefont {R.}~\bibnamefont {{Spann}}},\ }\href {\doibase
  10.1142/S2251171719500041} {\bibfield  {journal} {\bibinfo  {journal}
  {Journal of Astronomical Instrumentation}\ }\textbf {\bibinfo {volume} {8}},\
  \bibinfo {eid} {1950004} (\bibinfo {year} {2019})},\ \Eprint
  {http://arxiv.org/abs/1806.09531} {arXiv:1806.09531 [astro-ph.IM]}
  \BibitemShut {NoStop}%
\bibitem [{\citenamefont {{Burns}}\ \emph {et~al.}(2021)\citenamefont
  {{Burns}}, \citenamefont {{Bale}}, \citenamefont {{Bradley}}, \citenamefont
  {{Ahmed}}, \citenamefont {{Allen}}, \citenamefont {{Bowman}}, \citenamefont
  {{Furlanetto}}, \citenamefont {{MacDowall}}, \citenamefont {{Mirocha}},
  \citenamefont {{Nhan}}, \citenamefont {{Pivovaroff}}, \citenamefont
  {{Pulupa}}, \citenamefont {{Rapetti}}, \citenamefont {{Slosar}},\ and\
  \citenamefont {{Tauscher}}}]{DAPPER2021}%
  \BibitemOpen
  \bibfield  {author} {\bibinfo {author} {\bibfnamefont {J.}~\bibnamefont
  {{Burns}}}, \bibinfo {author} {\bibfnamefont {S.}~\bibnamefont {{Bale}}},
  \bibinfo {author} {\bibfnamefont {R.}~\bibnamefont {{Bradley}}}, \bibinfo
  {author} {\bibfnamefont {Z.}~\bibnamefont {{Ahmed}}}, \bibinfo {author}
  {\bibfnamefont {S.~W.}\ \bibnamefont {{Allen}}}, \bibinfo {author}
  {\bibfnamefont {J.}~\bibnamefont {{Bowman}}}, \bibinfo {author}
  {\bibfnamefont {S.}~\bibnamefont {{Furlanetto}}}, \bibinfo {author}
  {\bibfnamefont {R.}~\bibnamefont {{MacDowall}}}, \bibinfo {author}
  {\bibfnamefont {J.}~\bibnamefont {{Mirocha}}}, \bibinfo {author}
  {\bibfnamefont {B.}~\bibnamefont {{Nhan}}}, \bibinfo {author} {\bibfnamefont
  {M.}~\bibnamefont {{Pivovaroff}}}, \bibinfo {author} {\bibfnamefont
  {M.}~\bibnamefont {{Pulupa}}}, \bibinfo {author} {\bibfnamefont
  {D.}~\bibnamefont {{Rapetti}}}, \bibinfo {author} {\bibfnamefont
  {A.}~\bibnamefont {{Slosar}}}, \ and\ \bibinfo {author} {\bibfnamefont
  {K.}~\bibnamefont {{Tauscher}}},\ }\href@noop {} {\bibfield  {journal}
  {\bibinfo  {journal} {arXiv e-prints}\ ,\ \bibinfo {eid} {arXiv:2103.05085}}
  (\bibinfo {year} {2021})},\ \Eprint {http://arxiv.org/abs/2103.05085}
  {arXiv:2103.05085 [astro-ph.CO]} \BibitemShut {NoStop}%
\bibitem [{\citenamefont {DeBoer}\ \emph {et~al.}(2017)\citenamefont {DeBoer}
  \emph {et~al.}}]{DeBoer:2016tnn}%
  \BibitemOpen
  \bibfield  {author} {\bibinfo {author} {\bibfnamefont {D.~R.}\ \bibnamefont
  {DeBoer}} \emph {et~al.},\ }\href {\doibase 10.1088/1538-3873/129/974/045001}
  {\bibfield  {journal} {\bibinfo  {journal} {Publ. Astron. Soc. Pac.}\
  }\textbf {\bibinfo {volume} {129}},\ \bibinfo {pages} {045001} (\bibinfo
  {year} {2017})},\ \Eprint {http://arxiv.org/abs/1606.07473} {arXiv:1606.07473
  [astro-ph.IM]} \BibitemShut {NoStop}%
\bibitem [{\citenamefont {Bacon}\ \emph {et~al.}(2020)\citenamefont {Bacon}
  \emph {et~al.}}]{SKA:2018ckk}%
  \BibitemOpen
  \bibfield  {author} {\bibinfo {author} {\bibfnamefont {D.~J.}\ \bibnamefont
  {Bacon}} \emph {et~al.} (\bibinfo {collaboration} {SKA}),\ }\href {\doibase
  10.1017/pasa.2019.51} {\bibfield  {journal} {\bibinfo  {journal} {Publ.
  Astron. Soc. Austral.}\ }\textbf {\bibinfo {volume} {37}},\ \bibinfo {pages}
  {e007} (\bibinfo {year} {2020})},\ \Eprint {http://arxiv.org/abs/1811.02743}
  {arXiv:1811.02743 [astro-ph.CO]} \BibitemShut {NoStop}%
\bibitem [{\citenamefont {{Braun}}\ \emph {et~al.}(2019)\citenamefont
  {{Braun}}, \citenamefont {{Bonaldi}}, \citenamefont {{Bourke}}, \citenamefont
  {{Keane}},\ and\ \citenamefont {{Wagg}}}]{2019arXiv191212699B}%
  \BibitemOpen
  \bibfield  {author} {\bibinfo {author} {\bibfnamefont {R.}~\bibnamefont
  {{Braun}}}, \bibinfo {author} {\bibfnamefont {A.}~\bibnamefont {{Bonaldi}}},
  \bibinfo {author} {\bibfnamefont {T.}~\bibnamefont {{Bourke}}}, \bibinfo
  {author} {\bibfnamefont {E.}~\bibnamefont {{Keane}}}, \ and\ \bibinfo
  {author} {\bibfnamefont {J.}~\bibnamefont {{Wagg}}},\ }\href@noop {}
  {\bibfield  {journal} {\bibinfo  {journal} {arXiv e-prints}\ ,\ \bibinfo
  {eid} {arXiv:1912.12699}} (\bibinfo {year} {2019})},\ \Eprint
  {http://arxiv.org/abs/1912.12699} {arXiv:1912.12699 [astro-ph.IM]}
  \BibitemShut {NoStop}%
\bibitem [{\citenamefont {Mesinger}\ \emph {et~al.}(2011)\citenamefont
  {Mesinger}, \citenamefont {Furlanetto},\ and\ \citenamefont
  {Cen}}]{Mesinger:2010ne}%
  \BibitemOpen
  \bibfield  {author} {\bibinfo {author} {\bibfnamefont {A.}~\bibnamefont
  {Mesinger}}, \bibinfo {author} {\bibfnamefont {S.}~\bibnamefont
  {Furlanetto}}, \ and\ \bibinfo {author} {\bibfnamefont {R.}~\bibnamefont
  {Cen}},\ }\href {\doibase 10.1111/j.1365-2966.2010.17731.x} {\bibfield
  {journal} {\bibinfo  {journal} {Mon. Not. Roy. Astron. Soc.}\ }\textbf
  {\bibinfo {volume} {411}},\ \bibinfo {pages} {955} (\bibinfo {year}
  {2011})},\ \Eprint {http://arxiv.org/abs/1003.3878} {arXiv:1003.3878
  [astro-ph.CO]} \BibitemShut {NoStop}%
\bibitem [{\citenamefont {Greig}\ and\ \citenamefont
  {Mesinger}(2015)}]{Greig:2015qca}%
  \BibitemOpen
  \bibfield  {author} {\bibinfo {author} {\bibfnamefont {B.}~\bibnamefont
  {Greig}}\ and\ \bibinfo {author} {\bibfnamefont {A.}~\bibnamefont
  {Mesinger}},\ }\href {\doibase 10.1093/mnras/stv571} {\bibfield  {journal}
  {\bibinfo  {journal} {Mon. Not. Roy. Astron. Soc.}\ }\textbf {\bibinfo
  {volume} {449}},\ \bibinfo {pages} {4246} (\bibinfo {year} {2015})},\ \Eprint
  {http://arxiv.org/abs/1501.06576} {arXiv:1501.06576 [astro-ph.CO]}
  \BibitemShut {NoStop}%
\bibitem [{\citenamefont {Mu\~noz}\ \emph {et~al.}(2022)\citenamefont
  {Mu\~noz}, \citenamefont {Qin}, \citenamefont {Mesinger}, \citenamefont
  {Murray}, \citenamefont {Greig},\ and\ \citenamefont
  {Mason}}]{Munoz:2021psm}%
  \BibitemOpen
  \bibfield  {author} {\bibinfo {author} {\bibfnamefont {J.~B.}\ \bibnamefont
  {Mu\~noz}}, \bibinfo {author} {\bibfnamefont {Y.}~\bibnamefont {Qin}},
  \bibinfo {author} {\bibfnamefont {A.}~\bibnamefont {Mesinger}}, \bibinfo
  {author} {\bibfnamefont {S.~G.}\ \bibnamefont {Murray}}, \bibinfo {author}
  {\bibfnamefont {B.}~\bibnamefont {Greig}}, \ and\ \bibinfo {author}
  {\bibfnamefont {C.}~\bibnamefont {Mason}},\ }\href {\doibase
  10.1093/mnras/stac185} {\bibfield  {journal} {\bibinfo  {journal} {Mon. Not.
  Roy. Astron. Soc.}\ }\textbf {\bibinfo {volume} {511}},\ \bibinfo {pages}
  {3657} (\bibinfo {year} {2022})},\ \Eprint {http://arxiv.org/abs/2110.13919}
  {arXiv:2110.13919 [astro-ph.CO]} \BibitemShut {NoStop}%
\bibitem [{\citenamefont {Dalal}\ \emph {et~al.}(2010)\citenamefont {Dalal},
  \citenamefont {Pen},\ and\ \citenamefont {Seljak}}]{Dalal:2010yt}%
  \BibitemOpen
  \bibfield  {author} {\bibinfo {author} {\bibfnamefont {N.}~\bibnamefont
  {Dalal}}, \bibinfo {author} {\bibfnamefont {U.-L.}\ \bibnamefont {Pen}}, \
  and\ \bibinfo {author} {\bibfnamefont {U.}~\bibnamefont {Seljak}},\ }\href
  {\doibase 10.1088/1475-7516/2010/11/007} {\bibfield  {journal} {\bibinfo
  {journal} {JCAP}\ }\textbf {\bibinfo {volume} {1011}},\ \bibinfo {pages}
  {007} (\bibinfo {year} {2010})},\ \Eprint {http://arxiv.org/abs/1009.4704}
  {arXiv:1009.4704 [astro-ph.CO]} \BibitemShut {NoStop}%
\bibitem [{\citenamefont {Visbal}\ \emph {et~al.}(2012)\citenamefont {Visbal},
  \citenamefont {Barkana}, \citenamefont {Fialkov}, \citenamefont
  {Tseliakhovich},\ and\ \citenamefont {Hirata}}]{Visbal:2012aw}%
  \BibitemOpen
  \bibfield  {author} {\bibinfo {author} {\bibfnamefont {E.}~\bibnamefont
  {Visbal}}, \bibinfo {author} {\bibfnamefont {R.}~\bibnamefont {Barkana}},
  \bibinfo {author} {\bibfnamefont {A.}~\bibnamefont {Fialkov}}, \bibinfo
  {author} {\bibfnamefont {D.}~\bibnamefont {Tseliakhovich}}, \ and\ \bibinfo
  {author} {\bibfnamefont {C.}~\bibnamefont {Hirata}},\ }\href {\doibase
  10.1038/nature11177} {\bibfield  {journal} {\bibinfo  {journal} {Nature}\
  }\textbf {\bibinfo {volume} {487}},\ \bibinfo {pages} {70} (\bibinfo {year}
  {2012})},\ \Eprint {http://arxiv.org/abs/1201.1005} {arXiv:1201.1005
  [astro-ph.CO]} \BibitemShut {NoStop}%
\bibitem [{\citenamefont {McQuinn}\ and\ \citenamefont
  {O'Leary}(2012)}]{McQuinn:2012rt}%
  \BibitemOpen
  \bibfield  {author} {\bibinfo {author} {\bibfnamefont {M.}~\bibnamefont
  {McQuinn}}\ and\ \bibinfo {author} {\bibfnamefont {R.~M.}\ \bibnamefont
  {O'Leary}},\ }\href {\doibase 10.1088/0004-637X/760/1/3} {\bibfield
  {journal} {\bibinfo  {journal} {Astrophys. J.}\ }\textbf {\bibinfo {volume}
  {760}},\ \bibinfo {pages} {3} (\bibinfo {year} {2012})},\ \Eprint
  {http://arxiv.org/abs/1204.1345} {arXiv:1204.1345 [astro-ph.CO]} \BibitemShut
  {NoStop}%
\bibitem [{\citenamefont {Pritchard}\ and\ \citenamefont
  {Furlanetto}(2007)}]{Pritchard:2006sq}%
  \BibitemOpen
  \bibfield  {author} {\bibinfo {author} {\bibfnamefont {J.~R.}\ \bibnamefont
  {Pritchard}}\ and\ \bibinfo {author} {\bibfnamefont {S.~R.}\ \bibnamefont
  {Furlanetto}},\ }\href {\doibase 10.1111/j.1365-2966.2007.11519.x} {\bibfield
   {journal} {\bibinfo  {journal} {Mon. Not. Roy. Astron. Soc.}\ }\textbf
  {\bibinfo {volume} {376}},\ \bibinfo {pages} {1680} (\bibinfo {year}
  {2007})},\ \Eprint {http://arxiv.org/abs/astro-ph/0607234}
  {arXiv:astro-ph/0607234} \BibitemShut {NoStop}%
\bibitem [{\citenamefont {Mesinger}\ \emph {et~al.}(2013)\citenamefont
  {Mesinger}, \citenamefont {Ferrara},\ and\ \citenamefont
  {Spiegel}}]{Mesinger:2012ys}%
  \BibitemOpen
  \bibfield  {author} {\bibinfo {author} {\bibfnamefont {A.}~\bibnamefont
  {Mesinger}}, \bibinfo {author} {\bibfnamefont {A.}~\bibnamefont {Ferrara}}, \
  and\ \bibinfo {author} {\bibfnamefont {D.~S.}\ \bibnamefont {Spiegel}},\
  }\href {\doibase 10.1093/mnras/stt198} {\bibfield  {journal} {\bibinfo
  {journal} {Mon. Not. Roy. Astron. Soc.}\ }\textbf {\bibinfo {volume} {431}},\
  \bibinfo {pages} {621} (\bibinfo {year} {2013})},\ \Eprint
  {http://arxiv.org/abs/1210.7319} {arXiv:1210.7319 [astro-ph.CO]} \BibitemShut
  {NoStop}%
\bibitem [{\citenamefont {Fialkov}\ \emph {et~al.}(2014)\citenamefont
  {Fialkov}, \citenamefont {Barkana}, \citenamefont {Pinhas},\ and\
  \citenamefont {Visbal}}]{Fialkov:2013uwm}%
  \BibitemOpen
  \bibfield  {author} {\bibinfo {author} {\bibfnamefont {A.}~\bibnamefont
  {Fialkov}}, \bibinfo {author} {\bibfnamefont {R.}~\bibnamefont {Barkana}},
  \bibinfo {author} {\bibfnamefont {A.}~\bibnamefont {Pinhas}}, \ and\ \bibinfo
  {author} {\bibfnamefont {E.}~\bibnamefont {Visbal}},\ }\href {\doibase
  10.1093/mnrasl/slt135} {\bibfield  {journal} {\bibinfo  {journal} {Mon. Not.
  Roy. Astron. Soc.}\ }\textbf {\bibinfo {volume} {437}},\ \bibinfo {pages}
  {36} (\bibinfo {year} {2014})},\ \Eprint {http://arxiv.org/abs/1306.2354}
  {arXiv:1306.2354 [astro-ph.CO]} \BibitemShut {NoStop}%
\bibitem [{\citenamefont {Wise}\ and\ \citenamefont
  {Abel}(2008)}]{Wise:2007nb}%
  \BibitemOpen
  \bibfield  {author} {\bibinfo {author} {\bibfnamefont {J.~H.}\ \bibnamefont
  {Wise}}\ and\ \bibinfo {author} {\bibfnamefont {T.}~\bibnamefont {Abel}},\
  }\href {\doibase 10.1086/590417} {\bibfield  {journal} {\bibinfo  {journal}
  {Astrophys. J.}\ }\textbf {\bibinfo {volume} {685}},\ \bibinfo {pages} {40}
  (\bibinfo {year} {2008})},\ \Eprint {http://arxiv.org/abs/0710.3160}
  {arXiv:0710.3160 [astro-ph]} \BibitemShut {NoStop}%
\bibitem [{\citenamefont {Ali-Ha\"\i{}moud}\ \emph {et~al.}(2014)\citenamefont
  {Ali-Ha\"\i{}moud}, \citenamefont {Meerburg},\ and\ \citenamefont
  {Yuan}}]{Ali-Haimoud:2013hpa}%
  \BibitemOpen
  \bibfield  {author} {\bibinfo {author} {\bibfnamefont {Y.}~\bibnamefont
  {Ali-Ha\"\i{}moud}}, \bibinfo {author} {\bibfnamefont {P.~D.}\ \bibnamefont
  {Meerburg}}, \ and\ \bibinfo {author} {\bibfnamefont {S.}~\bibnamefont
  {Yuan}},\ }\href {\doibase 10.1103/PhysRevD.89.083506} {\bibfield  {journal}
  {\bibinfo  {journal} {Phys. Rev. D}\ }\textbf {\bibinfo {volume} {89}},\
  \bibinfo {pages} {083506} (\bibinfo {year} {2014})},\ \Eprint
  {http://arxiv.org/abs/1312.4948} {arXiv:1312.4948 [astro-ph.CO]} \BibitemShut
  {NoStop}%
\bibitem [{\citenamefont {Blas}\ \emph {et~al.}(2011)\citenamefont {Blas},
  \citenamefont {Lesgourgues},\ and\ \citenamefont {Tram}}]{Blas:2011rf}%
  \BibitemOpen
  \bibfield  {author} {\bibinfo {author} {\bibfnamefont {D.}~\bibnamefont
  {Blas}}, \bibinfo {author} {\bibfnamefont {J.}~\bibnamefont {Lesgourgues}}, \
  and\ \bibinfo {author} {\bibfnamefont {T.}~\bibnamefont {Tram}},\ }\href
  {\doibase 10.1088/1475-7516/2011/07/034} {\bibfield  {journal} {\bibinfo
  {journal} {JCAP}\ }\textbf {\bibinfo {volume} {07}},\ \bibinfo {pages} {034}
  (\bibinfo {year} {2011})},\ \Eprint {http://arxiv.org/abs/1104.2933}
  {arXiv:1104.2933 [astro-ph.CO]} \BibitemShut {NoStop}%
\bibitem [{\citenamefont {Ali-Haimoud}\ and\ \citenamefont
  {Hirata}(2010)}]{Ali-Haimoud:2010tlj}%
  \BibitemOpen
  \bibfield  {author} {\bibinfo {author} {\bibfnamefont {Y.}~\bibnamefont
  {Ali-Haimoud}}\ and\ \bibinfo {author} {\bibfnamefont {C.~M.}\ \bibnamefont
  {Hirata}},\ }\href {\doibase 10.1103/PhysRevD.82.063521} {\bibfield
  {journal} {\bibinfo  {journal} {Phys. Rev. D}\ }\textbf {\bibinfo {volume}
  {82}},\ \bibinfo {pages} {063521} (\bibinfo {year} {2010})},\ \Eprint
  {http://arxiv.org/abs/1006.1355} {arXiv:1006.1355 [astro-ph.CO]} \BibitemShut
  {NoStop}%
\bibitem [{\citenamefont {Ali-Haimoud}\ and\ \citenamefont
  {Hirata}(2011)}]{Ali-Haimoud:2010hou}%
  \BibitemOpen
  \bibfield  {author} {\bibinfo {author} {\bibfnamefont {Y.}~\bibnamefont
  {Ali-Haimoud}}\ and\ \bibinfo {author} {\bibfnamefont {C.~M.}\ \bibnamefont
  {Hirata}},\ }\href {\doibase 10.1103/PhysRevD.83.043513} {\bibfield
  {journal} {\bibinfo  {journal} {Phys. Rev. D}\ }\textbf {\bibinfo {volume}
  {83}},\ \bibinfo {pages} {043513} (\bibinfo {year} {2011})},\ \Eprint
  {http://arxiv.org/abs/1011.3758} {arXiv:1011.3758 [astro-ph.CO]} \BibitemShut
  {NoStop}%
\bibitem [{\citenamefont {Lee}\ and\ \citenamefont
  {Ali-Ha\"\i{}moud}(2020)}]{Lee:2020obi}%
  \BibitemOpen
  \bibfield  {author} {\bibinfo {author} {\bibfnamefont {N.}~\bibnamefont
  {Lee}}\ and\ \bibinfo {author} {\bibfnamefont {Y.}~\bibnamefont
  {Ali-Ha\"\i{}moud}},\ }\href {\doibase 10.1103/PhysRevD.102.083517}
  {\bibfield  {journal} {\bibinfo  {journal} {Phys. Rev. D}\ }\textbf {\bibinfo
  {volume} {102}},\ \bibinfo {pages} {083517} (\bibinfo {year} {2020})},\
  \Eprint {http://arxiv.org/abs/2007.14114} {arXiv:2007.14114 [astro-ph.CO]}
  \BibitemShut {NoStop}%
\bibitem [{\citenamefont {Hotinli}\ \emph
  {et~al.}(2021{\natexlab{b}})\citenamefont {Hotinli}, \citenamefont {Smith},
  \citenamefont {Madhavacheril},\ and\ \citenamefont
  {Kamionkowski}}]{Hotinli:2021hih}%
  \BibitemOpen
  \bibfield  {author} {\bibinfo {author} {\bibfnamefont {S.~C.}\ \bibnamefont
  {Hotinli}}, \bibinfo {author} {\bibfnamefont {K.~M.}\ \bibnamefont {Smith}},
  \bibinfo {author} {\bibfnamefont {M.~S.}\ \bibnamefont {Madhavacheril}}, \
  and\ \bibinfo {author} {\bibfnamefont {M.}~\bibnamefont {Kamionkowski}},\
  }\href {\doibase 10.1103/PhysRevD.104.083529} {\bibfield  {journal} {\bibinfo
   {journal} {Phys. Rev. D}\ }\textbf {\bibinfo {volume} {104}},\ \bibinfo
  {pages} {083529} (\bibinfo {year} {2021}{\natexlab{b}})},\ \Eprint
  {http://arxiv.org/abs/2108.02207} {arXiv:2108.02207 [astro-ph.CO]}
  \BibitemShut {NoStop}%
\bibitem [{\citenamefont {Hotinli}\ \emph
  {et~al.}(2022{\natexlab{b}})\citenamefont {Hotinli}, \citenamefont {Meyers},
  \citenamefont {Trendafilova}, \citenamefont {Green},\ and\ \citenamefont {van
  Engelen}}]{Hotinli:2021umk}%
  \BibitemOpen
  \bibfield  {author} {\bibinfo {author} {\bibfnamefont {S.~C.}\ \bibnamefont
  {Hotinli}}, \bibinfo {author} {\bibfnamefont {J.}~\bibnamefont {Meyers}},
  \bibinfo {author} {\bibfnamefont {C.}~\bibnamefont {Trendafilova}}, \bibinfo
  {author} {\bibfnamefont {D.}~\bibnamefont {Green}}, \ and\ \bibinfo {author}
  {\bibfnamefont {A.}~\bibnamefont {van Engelen}},\ }\href {\doibase
  10.1088/1475-7516/2022/04/020} {\bibfield  {journal} {\bibinfo  {journal}
  {JCAP}\ }\textbf {\bibinfo {volume} {04}},\ \bibinfo {pages} {020} (\bibinfo
  {year} {2022}{\natexlab{b}})},\ \Eprint {http://arxiv.org/abs/2111.15036}
  {arXiv:2111.15036 [astro-ph.CO]} \BibitemShut {NoStop}%
\bibitem [{\citenamefont {Zahn}\ and\ \citenamefont
  {Zaldarriaga}(2006)}]{Zahn:2005ap}%
  \BibitemOpen
  \bibfield  {author} {\bibinfo {author} {\bibfnamefont {O.}~\bibnamefont
  {Zahn}}\ and\ \bibinfo {author} {\bibfnamefont {M.}~\bibnamefont
  {Zaldarriaga}},\ }\href {\doibase 10.1086/508916} {\bibfield  {journal}
  {\bibinfo  {journal} {Astrophys. J.}\ }\textbf {\bibinfo {volume} {653}},\
  \bibinfo {pages} {922} (\bibinfo {year} {2006})},\ \Eprint
  {http://arxiv.org/abs/astro-ph/0511547} {arXiv:astro-ph/0511547} \BibitemShut
  {NoStop}%
\bibitem [{\citenamefont {Liu}\ \emph {et~al.}(2014{\natexlab{a}})\citenamefont
  {Liu}, \citenamefont {Parsons},\ and\ \citenamefont {Trott}}]{Liu:2014bba}%
  \BibitemOpen
  \bibfield  {author} {\bibinfo {author} {\bibfnamefont {A.}~\bibnamefont
  {Liu}}, \bibinfo {author} {\bibfnamefont {A.~R.}\ \bibnamefont {Parsons}}, \
  and\ \bibinfo {author} {\bibfnamefont {C.~M.}\ \bibnamefont {Trott}},\ }\href
  {\doibase 10.1103/PhysRevD.90.023018} {\bibfield  {journal} {\bibinfo
  {journal} {Phys. Rev. D}\ }\textbf {\bibinfo {volume} {90}},\ \bibinfo
  {pages} {023018} (\bibinfo {year} {2014}{\natexlab{a}})},\ \Eprint
  {http://arxiv.org/abs/1404.2596} {arXiv:1404.2596 [astro-ph.CO]} \BibitemShut
  {NoStop}%
\bibitem [{\citenamefont {Liu}\ \emph {et~al.}(2014{\natexlab{b}})\citenamefont
  {Liu}, \citenamefont {Parsons},\ and\ \citenamefont {Trott}}]{Liu:2014yxa}%
  \BibitemOpen
  \bibfield  {author} {\bibinfo {author} {\bibfnamefont {A.}~\bibnamefont
  {Liu}}, \bibinfo {author} {\bibfnamefont {A.~R.}\ \bibnamefont {Parsons}}, \
  and\ \bibinfo {author} {\bibfnamefont {C.~M.}\ \bibnamefont {Trott}},\ }\href
  {\doibase 10.1103/PhysRevD.90.023019} {\bibfield  {journal} {\bibinfo
  {journal} {Phys. Rev. D}\ }\textbf {\bibinfo {volume} {90}},\ \bibinfo
  {pages} {023019} (\bibinfo {year} {2014}{\natexlab{b}})},\ \Eprint
  {http://arxiv.org/abs/1404.4372} {arXiv:1404.4372 [astro-ph.CO]} \BibitemShut
  {NoStop}%
\bibitem [{\citenamefont {Romeo}\ \emph {et~al.}(2018)\citenamefont {Romeo},
  \citenamefont {Metcalf},\ and\ \citenamefont {Pourtsidou}}]{Romeo:2017zwt}%
  \BibitemOpen
  \bibfield  {author} {\bibinfo {author} {\bibfnamefont {A.}~\bibnamefont
  {Romeo}}, \bibinfo {author} {\bibfnamefont {R.~B.}\ \bibnamefont {Metcalf}},
  \ and\ \bibinfo {author} {\bibfnamefont {A.}~\bibnamefont {Pourtsidou}},\
  }\href {\doibase 10.1093/mnras/stx2733} {\bibfield  {journal} {\bibinfo
  {journal} {Mon. Not. Roy. Astron. Soc.}\ }\textbf {\bibinfo {volume} {474}},\
  \bibinfo {pages} {1787} (\bibinfo {year} {2018})},\ \Eprint
  {http://arxiv.org/abs/1708.01235} {arXiv:1708.01235 [astro-ph.CO]}
  \BibitemShut {NoStop}%
\bibitem [{\citenamefont {Foreman}\ \emph {et~al.}(2018)\citenamefont
  {Foreman}, \citenamefont {Meerburg}, \citenamefont {van Engelen},\ and\
  \citenamefont {Meyers}}]{Foreman:2018gnv}%
  \BibitemOpen
  \bibfield  {author} {\bibinfo {author} {\bibfnamefont {S.}~\bibnamefont
  {Foreman}}, \bibinfo {author} {\bibfnamefont {P.~D.}\ \bibnamefont
  {Meerburg}}, \bibinfo {author} {\bibfnamefont {A.}~\bibnamefont {van
  Engelen}}, \ and\ \bibinfo {author} {\bibfnamefont {J.}~\bibnamefont
  {Meyers}},\ }\href {\doibase 10.1088/1475-7516/2018/07/046} {\bibfield
  {journal} {\bibinfo  {journal} {JCAP}\ }\textbf {\bibinfo {volume} {07}},\
  \bibinfo {pages} {046} (\bibinfo {year} {2018})},\ \Eprint
  {http://arxiv.org/abs/1803.04975} {arXiv:1803.04975 [astro-ph.CO]}
  \BibitemShut {NoStop}%
\bibitem [{\citenamefont {Aghanim}\ \emph {et~al.}(2020)\citenamefont {Aghanim}
  \emph {et~al.}}]{Planck:2018vyg}%
  \BibitemOpen
  \bibfield  {author} {\bibinfo {author} {\bibfnamefont {N.}~\bibnamefont
  {Aghanim}} \emph {et~al.} (\bibinfo {collaboration} {Planck}),\ }\href
  {\doibase 10.1051/0004-6361/201833910} {\bibfield  {journal} {\bibinfo
  {journal} {Astron. Astrophys.}\ }\textbf {\bibinfo {volume} {641}},\ \bibinfo
  {pages} {A6} (\bibinfo {year} {2020})},\ \bibinfo {note} {[Erratum:
  Astron.Astrophys. 652, C4 (2021)]},\ \Eprint
  {http://arxiv.org/abs/1807.06209} {arXiv:1807.06209 [astro-ph.CO]}
  \BibitemShut {NoStop}%
\bibitem [{\citenamefont {Gagnon-Hartman}\ \emph {et~al.}(2021)\citenamefont
  {Gagnon-Hartman}, \citenamefont {Cui}, \citenamefont {Kennedy}, \citenamefont
  {Liu},\ and\ \citenamefont {Ravanbakhsh}}]{Gagnon-Hartman:2021erd}%
  \BibitemOpen
  \bibfield  {author} {\bibinfo {author} {\bibfnamefont {S.}~\bibnamefont
  {Gagnon-Hartman}}, \bibinfo {author} {\bibfnamefont {Y.}~\bibnamefont {Cui}},
  \bibinfo {author} {\bibfnamefont {J.}~\bibnamefont {Kennedy}}, \bibinfo
  {author} {\bibfnamefont {A.}~\bibnamefont {Liu}}, \ and\ \bibinfo {author}
  {\bibfnamefont {S.}~\bibnamefont {Ravanbakhsh}},\ }\href {\doibase
  10.1093/mnras/stab1158} {\bibfield  {journal} {\bibinfo  {journal} {Mon. Not.
  Roy. Astron. Soc.}\ }\textbf {\bibinfo {volume} {504}},\ \bibinfo {pages}
  {4716} (\bibinfo {year} {2021})},\ \Eprint {http://arxiv.org/abs/2102.08382}
  {arXiv:2102.08382 [astro-ph.CO]} \BibitemShut {NoStop}%
\bibitem [{\citenamefont {Beane}\ \emph {et~al.}(2019)\citenamefont {Beane},
  \citenamefont {Villaescusa-Navarro},\ and\ \citenamefont
  {Lidz}}]{Beane:2018dzk}%
  \BibitemOpen
  \bibfield  {author} {\bibinfo {author} {\bibfnamefont {A.}~\bibnamefont
  {Beane}}, \bibinfo {author} {\bibfnamefont {F.}~\bibnamefont
  {Villaescusa-Navarro}}, \ and\ \bibinfo {author} {\bibfnamefont
  {A.}~\bibnamefont {Lidz}},\ }\href {\doibase 10.3847/1538-4357/ab0a08}
  {\bibfield  {journal} {\bibinfo  {journal} {Astrophys. J.}\ }\textbf
  {\bibinfo {volume} {874}},\ \bibinfo {pages} {133} (\bibinfo {year}
  {2019})},\ \Eprint {http://arxiv.org/abs/1811.10609} {arXiv:1811.10609
  [astro-ph.CO]} \BibitemShut {NoStop}%
\bibitem [{\citenamefont {Li}\ \emph {et~al.}(2019)\citenamefont {Li},
  \citenamefont {Xu}, \citenamefont {Ma}, \citenamefont {Zhu}, \citenamefont
  {Hu}, \citenamefont {Zhu}, \citenamefont {Gu}, \citenamefont {Shan},
  \citenamefont {Zhu},\ and\ \citenamefont {Wu}}]{Li:2019znt}%
  \BibitemOpen
  \bibfield  {author} {\bibinfo {author} {\bibfnamefont {W.}~\bibnamefont
  {Li}}, \bibinfo {author} {\bibfnamefont {H.}~\bibnamefont {Xu}}, \bibinfo
  {author} {\bibfnamefont {Z.}~\bibnamefont {Ma}}, \bibinfo {author}
  {\bibfnamefont {R.}~\bibnamefont {Zhu}}, \bibinfo {author} {\bibfnamefont
  {D.}~\bibnamefont {Hu}}, \bibinfo {author} {\bibfnamefont {Z.}~\bibnamefont
  {Zhu}}, \bibinfo {author} {\bibfnamefont {J.}~\bibnamefont {Gu}}, \bibinfo
  {author} {\bibfnamefont {C.}~\bibnamefont {Shan}}, \bibinfo {author}
  {\bibfnamefont {J.}~\bibnamefont {Zhu}}, \ and\ \bibinfo {author}
  {\bibfnamefont {X.-P.}\ \bibnamefont {Wu}},\ }\href {\doibase
  10.1093/mnras/stz582} {\bibfield  {journal} {\bibinfo  {journal} {Mon. Not.
  Roy. Astron. Soc.}\ }\textbf {\bibinfo {volume} {485}},\ \bibinfo {pages}
  {2628} (\bibinfo {year} {2019})},\ \Eprint {http://arxiv.org/abs/1902.09278}
  {arXiv:1902.09278 [astro-ph.IM]} \BibitemShut {NoStop}%
\bibitem [{\citenamefont {Villanueva-Domingo}\ and\ \citenamefont
  {Villaescusa-Navarro}(2021)}]{Villanueva-Domingo:2020wpt}%
  \BibitemOpen
  \bibfield  {author} {\bibinfo {author} {\bibfnamefont {P.}~\bibnamefont
  {Villanueva-Domingo}}\ and\ \bibinfo {author} {\bibfnamefont
  {F.}~\bibnamefont {Villaescusa-Navarro}},\ }\href {\doibase
  10.3847/1538-4357/abd245} {\bibfield  {journal} {\bibinfo  {journal}
  {Astrophys. J.}\ }\textbf {\bibinfo {volume} {907}},\ \bibinfo {pages} {44}
  (\bibinfo {year} {2021})},\ \Eprint {http://arxiv.org/abs/2006.14305}
  {arXiv:2006.14305 [astro-ph.CO]} \BibitemShut {NoStop}%
\bibitem [{\citenamefont {Makinen}\ \emph {et~al.}(2021)\citenamefont
  {Makinen}, \citenamefont {Lancaster}, \citenamefont {Villaescusa-Navarro},
  \citenamefont {Melchior}, \citenamefont {Ho}, \citenamefont
  {Perreault-Levasseur},\ and\ \citenamefont {Spergel}}]{Makinen:2020gvh}%
  \BibitemOpen
  \bibfield  {author} {\bibinfo {author} {\bibfnamefont {T.~L.}\ \bibnamefont
  {Makinen}}, \bibinfo {author} {\bibfnamefont {L.}~\bibnamefont {Lancaster}},
  \bibinfo {author} {\bibfnamefont {F.}~\bibnamefont {Villaescusa-Navarro}},
  \bibinfo {author} {\bibfnamefont {P.}~\bibnamefont {Melchior}}, \bibinfo
  {author} {\bibfnamefont {S.}~\bibnamefont {Ho}}, \bibinfo {author}
  {\bibfnamefont {L.}~\bibnamefont {Perreault-Levasseur}}, \ and\ \bibinfo
  {author} {\bibfnamefont {D.~N.}\ \bibnamefont {Spergel}},\ }\href {\doibase
  10.1088/1475-7516/2021/04/081} {\bibfield  {journal} {\bibinfo  {journal}
  {JCAP}\ }\textbf {\bibinfo {volume} {04}},\ \bibinfo {pages} {081} (\bibinfo
  {year} {2021})},\ \Eprint {http://arxiv.org/abs/2010.15843} {arXiv:2010.15843
  [astro-ph.CO]} \BibitemShut {NoStop}%
\bibitem [{\citenamefont {Bharadwaj}\ and\ \citenamefont
  {Ali}(2004)}]{Bharadwaj:2004nr}%
  \BibitemOpen
  \bibfield  {author} {\bibinfo {author} {\bibfnamefont {S.}~\bibnamefont
  {Bharadwaj}}\ and\ \bibinfo {author} {\bibfnamefont {S.~S.}\ \bibnamefont
  {Ali}},\ }\href {\doibase 10.1111/j.1365-2966.2004.07907.x} {\bibfield
  {journal} {\bibinfo  {journal} {Mon. Not. Roy. Astron. Soc.}\ }\textbf
  {\bibinfo {volume} {352}},\ \bibinfo {pages} {142} (\bibinfo {year}
  {2004})},\ \Eprint {http://arxiv.org/abs/astro-ph/0401206}
  {arXiv:astro-ph/0401206} \BibitemShut {NoStop}%
\bibitem [{\citenamefont {Barkana}\ and\ \citenamefont
  {Loeb}(2005)}]{Barkana:2004zy}%
  \BibitemOpen
  \bibfield  {author} {\bibinfo {author} {\bibfnamefont {R.}~\bibnamefont
  {Barkana}}\ and\ \bibinfo {author} {\bibfnamefont {A.}~\bibnamefont {Loeb}},\
  }\href {\doibase 10.1086/430599} {\bibfield  {journal} {\bibinfo  {journal}
  {Astrophys. J. Lett.}\ }\textbf {\bibinfo {volume} {624}},\ \bibinfo {pages}
  {L65} (\bibinfo {year} {2005})},\ \Eprint
  {http://arxiv.org/abs/astro-ph/0409572} {arXiv:astro-ph/0409572} \BibitemShut
  {NoStop}%
\bibitem [{\citenamefont {Mellema}\ \emph {et~al.}(2013)\citenamefont {Mellema}
  \emph {et~al.}}]{Mellema:2012ht}%
  \BibitemOpen
  \bibfield  {author} {\bibinfo {author} {\bibfnamefont {G.}~\bibnamefont
  {Mellema}} \emph {et~al.},\ }\href {\doibase 10.1007/s10686-013-9334-5}
  {\bibfield  {journal} {\bibinfo  {journal} {Exper. Astron.}\ }\textbf
  {\bibinfo {volume} {36}},\ \bibinfo {pages} {235} (\bibinfo {year} {2013})},\
  \Eprint {http://arxiv.org/abs/1210.0197} {arXiv:1210.0197 [astro-ph.CO]}
  \BibitemShut {NoStop}%
\bibitem [{\citenamefont {Mu\~noz}(2023)}]{Munoz:2023kkg}%
  \BibitemOpen
  \bibfield  {author} {\bibinfo {author} {\bibfnamefont {J.~B.}\ \bibnamefont
  {Mu\~noz}},\ }\href {\doibase 10.1093/mnras/stad1512} {\  (\bibinfo {year}
  {2023}),\ 10.1093/mnras/stad1512},\ \Eprint {http://arxiv.org/abs/2302.08506}
  {arXiv:2302.08506 [astro-ph.CO]} \BibitemShut {NoStop}%
\bibitem [{\citenamefont {Flitter}\ and\ \citenamefont
  {Kovetz}(2023{\natexlab{a}})}]{Flitter:2023mjj}%
  \BibitemOpen
  \bibfield  {author} {\bibinfo {author} {\bibfnamefont {J.}~\bibnamefont
  {Flitter}}\ and\ \bibinfo {author} {\bibfnamefont {E.~D.}\ \bibnamefont
  {Kovetz}},\ }\href@noop {} {\  (\bibinfo {year} {2023}{\natexlab{a}})},\
  \Eprint {http://arxiv.org/abs/2309.03942} {arXiv:2309.03942 [astro-ph.CO]}
  \BibitemShut {NoStop}%
\bibitem [{\citenamefont {Flitter}\ and\ \citenamefont
  {Kovetz}(2023{\natexlab{b}})}]{Flitter:2023rzv}%
  \BibitemOpen
  \bibfield  {author} {\bibinfo {author} {\bibfnamefont {J.}~\bibnamefont
  {Flitter}}\ and\ \bibinfo {author} {\bibfnamefont {E.~D.}\ \bibnamefont
  {Kovetz}},\ }\href@noop {} {\  (\bibinfo {year} {2023}{\natexlab{b}})},\
  \Eprint {http://arxiv.org/abs/2309.03948} {arXiv:2309.03948 [astro-ph.CO]}
  \BibitemShut {NoStop}%
\end{thebibliography}%

\end{document}